\documentclass[fleqn,usenatbib]{mnras}

\usepackage{newtxtext,newtxmath}

\usepackage[T1]{fontenc}

\DeclareRobustCommand{\VAN}[3]{#2}
\let\VANthebibliography\thebibliography
\def\thebibliography{\DeclareRobustCommand{\VAN}[3]{##3}\VANthebibliography}

\usepackage{graphicx}	
\usepackage{amsmath}	
\usepackage{eso-pic}

\AddToShipoutPictureBG*{%
  \AtPageUpperLeft{%
    \hspace{0.75\paperwidth}%
    \raisebox{-3.5\baselineskip}{%
      \makebox[0pt][l]{\textnormal{DES-2020-0537}}
}}}%

\AddToShipoutPictureBG*{%
  \AtPageUpperLeft{%
    \hspace{0.75\paperwidth}%
    \raisebox{-4.5\baselineskip}{%
      \makebox[0pt][l]{\textnormal{FERMILAB-PUB-20-474-AE}}
}}}%

\newcommand{\mstellar}{\ensuremath{\mathrm{M}_{\mathrm{stellar}}}}
\defcitealias{cosimo13}{I13}

\title{Understanding the extreme luminosity of DES14X2fna}

\author[DES Collaboration]{
\parbox{\textwidth}{
\Large
M.~Grayling,$^{1,2}$
C.~P.~Guti\'errez,$^{1}$
M.~Sullivan,$^{1}$
P.~Wiseman,$^{1}$
M.~Vincenzi,$^{3}$
S.~Gonz\'alez-Gait\'an,$^{4}$
B.~E.~Tucker,$^{5}$
L.~Galbany,$^{6}$
L.~Kelsey,$^{1}$
C.~Lidman,$^{5}$
E.~Swann,$^{3}$
D.~Carollo,$^{7}$
K.~Glazebrook,$^{8}$
G.~F.~Lewis,$^{9}$
A.~M\"oller,$^{10}$
S.~R.~Hinton,$^{11}$
M.~Smith,$^{1}$
S.~A.~Uddin,$^{12}$
T.~M.~C.~Abbott,$^{13}$
M.~Aguena,$^{14,15}$
S.~Avila,$^{16}$
E.~Bertin,$^{17,18}$
S.~Bhargava,$^{19}$
D.~Brooks,$^{20}$
A.~Carnero~Rosell,$^{21,22}$
M.~Carrasco~Kind,$^{23,24}$
J.~Carretero,$^{25}$
M.~Costanzi,$^{26,27}$
L.~N.~da Costa,$^{15,28}$
J.~De~Vicente,$^{29}$
S.~Desai,$^{30}$
H.~T.~Diehl,$^{31}$
P.~Doel,$^{20}$
S.~Everett,$^{32}$
I.~Ferrero,$^{33}$
P.~Fosalba,$^{34,35}$
J.~Frieman,$^{31,36}$
J.~Garc\'ia-Bellido,$^{16}$
E.~Gaztanaga,$^{34,35}$
D.~Gruen,$^{37,38,39}$
R.~A.~Gruendl,$^{23,24}$
J.~Gschwend,$^{15,28}$
G.~Gutierrez,$^{31}$
B.~Hoyle,$^{40,41,42}$
K.~Kuehn,$^{43,44}$
N.~Kuropatkin,$^{31}$
M.~Lima,$^{14,15}$
N.~MacCrann,$^{45,46}$
J.~L.~Marshall,$^{47}$
P.~Martini,$^{45,48,49}$
R.~Miquel,$^{25,50}$
R.~Morgan,$^{51}$
A.~Palmese,$^{31,36}$
F.~Paz-Chinch\'{o}n,$^{24,52}$
A.~A.~Plazas,$^{53}$
A.~K.~Romer,$^{19}$
C.~S{\'a}nchez,$^{54}$
E.~Sanchez,$^{29}$
V.~Scarpine,$^{31}$
S.~Serrano,$^{34,35}$
I.~Sevilla-Noarbe,$^{29}$
M.~Soares-Santos,$^{55}$
E.~Suchyta,$^{56}$
G.~Tarle,$^{55}$
D.~Thomas,$^{3}$
C.~To,$^{37,38,39}$
T.~N.~Varga,$^{41,42}$
A.~R.~Walker,$^{13}$
and R.D.~Wilkinson$^{19}$
\begin{center} (DES Collaboration) \end{center}
\textit{Affiliations are listed at end of paper}
}
}

\date{Accepted XXX. Received YYY; in original form ZZZ}

\pubyear{2020}

\begin{document}
\label{firstpage}
\pagerange{\pageref{firstpage}--\pageref{lastpage}}
\maketitle

\begin{abstract}
We present DES14X2fna, a high-luminosity, fast-declining type IIb supernova (SN IIb) at redshift $z=0.0453$, detected by the Dark Energy Survey (DES). DES14X2fna is an unusual member of its class, with a light curve showing a broad, luminous peak reaching $M_r\simeq-19.3$\,mag 20 days after explosion. This object does not show a linear decline tail in the light curve until $\simeq$60 days after explosion, after which it declines very rapidly (4.38$\pm$0.10 mag 100 d$^{-1}$ in $r$-band). By fitting semi-analytic models to the photometry of DES14X2fna, we find that its light curve cannot be explained by a standard $^{56}$Ni decay model as this is unable to fit the peak and fast tail decline observed. Inclusion of either interaction with surrounding circumstellar material or a rapidly-rotating neutron star (magnetar) significantly increases the quality of the model fit. We also investigate the possibility for an object similar to DES14X2fna to act as a contaminant in photometric samples of SNe Ia for cosmology, finding that a similar simulated object is misclassified by a recurrent neural network (RNN)-based photometric classifier as a SN Ia in $\sim$1.1-2.4 per cent of cases in DES, depending on the probability threshold used for a positive classification.

\end{abstract}




\section{Introduction}

Core-collapse supernovae (SNe) are a diverse and heterogeneous population of events, with the variety of observed sub-types reflecting the complexity of their possible progenitor systems and astrophysics. Type II SNe (SNe II) are events displaying hydrogen lines in their photospheric spectra, type Ib SNe (SNe Ib) lack hydrogen but do contain helium, while type Ic SNe (SNe Ic) lack both. Type IIb supernovae (SNe IIb) are an intermediate class, displaying hydrogen lines at early times before the appearance of helium lines as seen in SNe Ib \citep{Filippenko97,GalYam17,modjaz_review}.

The commonly-accepted physical explanation for SNe IIb is that their progenitors have had their outer hydrogen envelope partially, but not fully, stripped away. SNe Ib have this envelope fully stripped, leading to spectra with helium but not hydrogen, while the progenitors of SNe Ic are stripped of both hydrogen and helium. A further class of stripped-envelope SNe, SN Ic with broad lines (SN Ic-BLs), shows similar spectroscopic features to SNe Ic, but with  broader features indicating high expansion velocities and an energetic explosion.

The exact mechanism driving the envelope stripping of these SNe is still open for debate, but proposed solutions include stellar winds \citep{wind_stripping} and interaction with a binary companion \citep{binary_stripping} in the case of a binary progenitor system. Stellar winds require massive progenitors of $\gtrsim25$--30\,M$_{\sun}$ in order to remove at least the majority of the hydrogen envelope \citep{progenitor_mass}. Pre-explosion and late-time images of the SN environment of the well-studied SN IIb SN~1993J \citep{93J_progenitor_images,93J_binary} indicate the presence of a binary system, with evidence of a binary companion also found in SN~2001ig \citep{2001ig_binary} and SN~2011dh \citep{2011dh_binary}. However, deep imaging studies of the SN IIb remnant Cassiopeia A have not indicated a binary companion \citep{CasA1,CasA2}, suggesting that both of these progenitor scenarios may occur.

In the canonical picture of a SN IIb, the light curve is driven by the radioactive decay chain of $^{56}$Ni synthesised in the explosion, which subsequently decays into $^{56}$Co and then stable $^{56}$Fe. Treatments of this radioactive decay model, for example the commonly used \lq Arnett\rq\ model from \citet{arnett82} and more recently \citet{k&k}, allow for various properties of the explosion to be estimated. Some other types of core-collapse SNe are primarily driven by different physical processes (e.g. interaction with a surrounding circumstellar material (CSM) for SNe IIn; \citealt{IIn_CSM}), although a $^{56}$Ni decay model can still be used to estimate some explosion properties \citep[e.g.,][]{prentice_bol,mass_distribution}. For SNe with light curves driven by $^{56}$Ni decay such as SNe IIb, a more luminous SN indicates a higher synthesised mass of $^{56}$Ni to power the peak of the light curve. 

Energetic SNe Ic-BL, however, are not well fit by this model, which cannot reproduce both the luminous peaks and the late-time light curves of these objects. These objects have traditionally been fit with a two-component model, with the light curve peak and broad spectral features powered by a fast-moving component and the exponential decline powered by a slower-moving dense component \citep{two_component_model}. However, more recently magnetar models have also proved successful in fitting the light curves of these objects. In this scenario, the light curve is powered by a combination of radioactive decay and energy injected into the system by a central engine, the spin down of a rapidly rotating neutron star \citep{ magnetar1,magnetar2}. \citet{98bwmag} shows that a combination of a magnetar with $^{56}$Ni decay successfully fits the light curves of the SNe Ic-BL SN~1998bw and SN~2002ap, with the magnetar able to explain the deviation of the late-time light curve from the intermediate exponential decline. 

In addition to this, some SNe IIb (e.g. 1993J, \citealt{IIb_sample9}, 2016gkg, \citealt{IIb_sample32}) exhibit an initial peak in their light curves which has been attributed to post-shock-breakout cooling in the case of a progenitor with a compact core surrounded by extended, low-mass material \citep{shock_breakour_nakar}. This typically occurs over a short period of a few days, and is not observed in all SNe IIb (e.g. SN~2008ax;~\citealt{IIb_sample23}), potentially because the SN is not discovered until after this phase. This pre-max bump can help infer properties of the progenitor including radius and binarity using hydrodynamic simulations \citep[e.g.,][]{Bersten2012,Piro2015,Sapir2017}.

As the spectroscopic properties of a stripped-envelope SN differ primarily due to the degree of stripping of the progenitor star, it is an open question as to whether SNe IIb, Ib, Ic and Ic-BL are distinct classes or part of a continuum \citep{modjaz_review}. \citet{Galbany2018} finds that SNe IIb have unusual host properties compared with other core-collapse SN hosts, having particularly low metallicity and star formation rate (SFR). SNe Ib, Ic and Ic-BL have been previously observed with peak absolute magnitudes from -16 up to and even brighter than -20, as shown in Fig. 2 and 3 of \citet{modjaz_review}. Historically, SNe IIb have exhibited less diversity in peak luminosity, ranging from a peak $r$/$R$-band absolute magnitude of roughly -16.5 to -18. However, the recent discovery of ASASSN-18am with a peak M$_V$ $\sim$ -19.7 \citep{asassn} demonstrated that SNe IIb can reach considerable luminosities.


In this paper, we present photometry and spectroscopy of DES14X2fna, an unusual and very luminous SN IIb discovered by Dark Energy Survey (DES) under the Dark Energy Survey Supernova Programme \citep[DES-SN,][]{DES-SN} and exhibiting very different properties to those shown by previously observed SNe IIb. In section \ref{obs}, we detail our observations of DES14X2fna. In section \ref{analyse}, we analyse the spectroscopic and photometric properties of both the SN and its host, and compare to samples of historic SNe. We consider a variety of semi-analytic models to explain the luminosity and evolution of DES14X2fna in section \ref{mosfit}. Next, we discuss the possible mechanisms which could drive the unusual light curve of DES14X2fna and consider the possibility that a similar object could act as a contaminant in photometric samples of SNe Ia in section \ref{discuss}, before concluding in section \ref{conclusions}. Throughout this analysis, we have assumed a flat $\Lambda$CDM cosmology with $\Omega_M = 0.3$, $\Omega_\Lambda = 0.7$ and H$_0 = 70$ km s$^{-1}$ Mpc$^{-1}$.

\section{Observations}
\label{obs}

DES14X2fna was discovered by DES-SN in an $r$-band image captured by Dark Energy Camera \citep[DECam][]{DECam} at an apparent magnitude of $m_r=19.1$\,mag. This discovery was on 2014 October 1 (MJD 56931.2), with a previous non-detection on the 2014 September 24 (MJD 56924.2) at $m_z\sim23.7$. The transient was located in a faint host galaxy with $M_r\sim-16$ at $z=0.0453$ \footnote{Obtained from narrow host galaxy emission features}, at position $\alpha$ = 02$^\text{h}$23$^\text{m}$15\fs64, $\delta$ = -07\degr05\arcmin20\farcs8 (J2000). Based on the epochs of first detection and last non-detection, we adopt an explosion date of MJD $56927.7\pm3.5$\,d. 

After discovery, $griz$ photometric coverage was acquired by DES-SN until January 2015. Photometric measurements were made using the pipeline outlined in \citet{DES_pipeline1} and \citet{DES_pipeline2}, which uses template subtraction to remove the host galaxy contribution to the image using a point-spread-function (PSF) matching routine. From this difference image, PSF-fitting is used to measure the SN photometry. We correct the photometry for Milky Way extinction using dust maps from \citet{ext_map}, assuming $R_V = 3.1$. We assume negligible host galaxy extinction  - we verify this by comparing H$\alpha$ and H$\beta$ flux ratios in spectroscopy of the host, which we find to be consistent with E(B-V)$_\text{host}\sim0$ \citep{host_extinction}.

\begin{figure*}
\centering
\includegraphics[width = \textwidth]{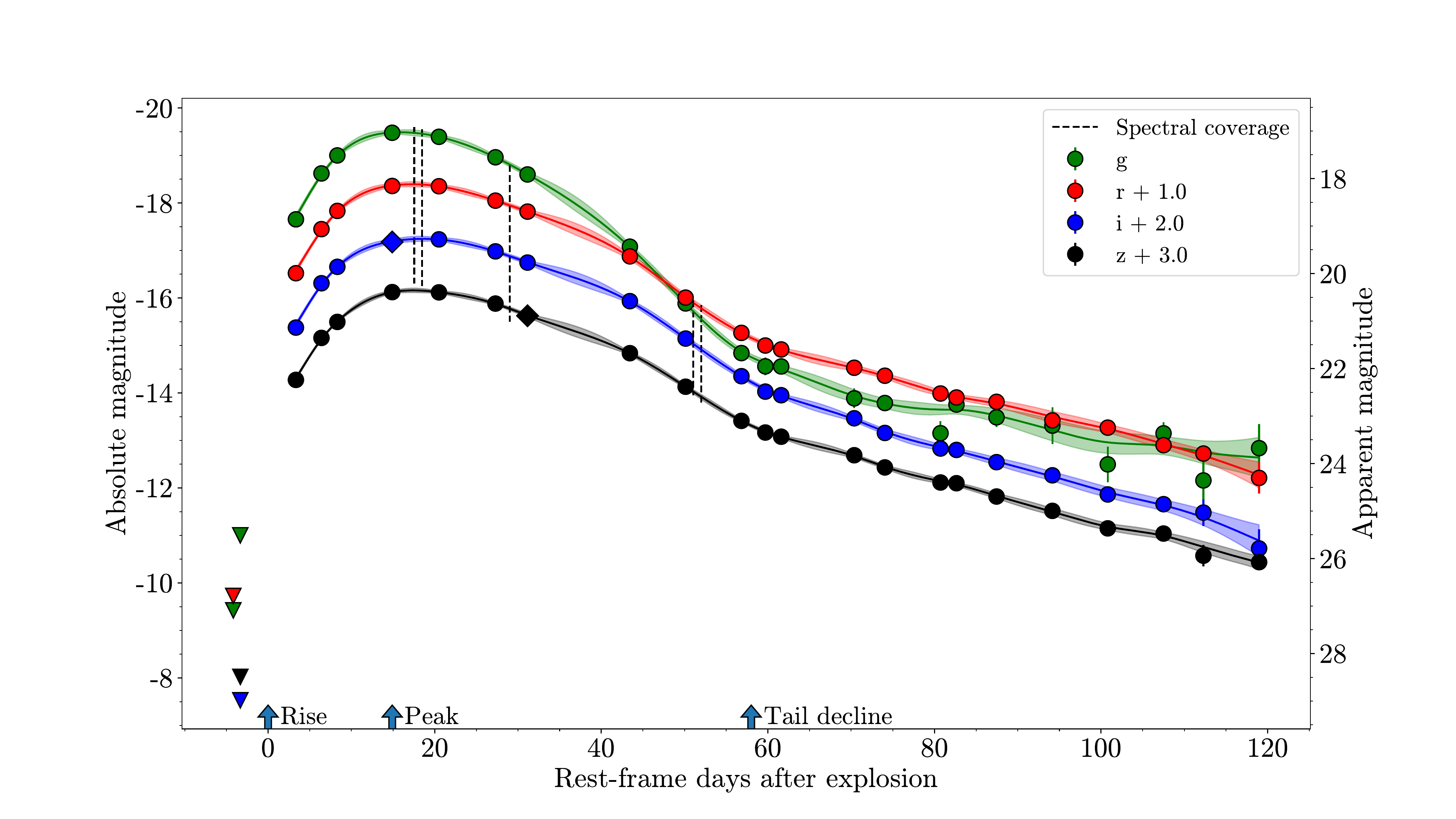}
\caption{$griz$ light curves of DES14X2fna, corrected for Milky Way extinction and k-corrected to the rest frame. Missing epochs of data have been interpolated using GP-interpolation. The dashed vertical lines indicate the epochs of spectral coverage, and the triangles prior to explosion denote upper limits. Diamond markers for data points indicate that they were reconstructed based on GP-interpolation of the full observer-frame light curve.}
\label{LC}
\end{figure*}

Photometric data was then K-corrected into the rest-frame. We do this using the SED templates of DES14X2fna from Hounsell et al. (in prep.), which we interpolate to epochs where we have observations and calibrate (\lq mangle\rq) to match our photometry. The mangling process required simultaneous observations in each photometric band - although the DES observations were near simultaneous across different bands, in a few instances data for a given band was missing. To complete our data and give fully simultaneous data, we interpolate the observed light curves using Gaussian Processes (GP; \citealt{GP_Rasmussen}). These were implemented using the \textsc{python} package \textsc{george} \citep{hodlr}, following the process outlined in \citet{slsne}. The observed photometry without any corrections is detailed in Table~\ref{phot_table}, and the corrected rest-frame light curves are shown in Fig.~\ref{LC} and also detailed in Table~\ref{phot_table}. The quoted uncertainties are purely statistical without incorporating any systematic uncertainties. Due to the high signal-to-noise observations of DES14X2fna, these uncertainties are very small and reach millimag levels at peak. In practice, the uncertainties will be larger than this - for our analysis, we add the statistical errors in quadrature with a value of 0.05 mag to represent systematic uncertainty. This value was selected as the smallest statistical error we could apply to obtain a stable GP fit across the full light curve with minimal unphysical undulations. This error is reflected in the rest-frame photometry in in Table~\ref{phot_table}. The GP-interpolated light curve used to reconstruct missing data was applied in the observer-frame - as such, it is not plotted with the rest-frame data in this figure. Instead, the GP-interpolation shown is obtained from this rest-frame data and is included to illustrate fits obtained from GP-interpolation. The length scale of the GP fit was determined by maximising the likelihood of the interpolation.

\begin{table*}
\scriptsize
\caption{Observed and rest-frame photometry of DES14X2fna, quoted in AB magnitudes in the natural DECam system. Note that this photometry has not been corrected for Milky Way extinction and quoted uncertainties are purely statistical. For our analysis, these have been added in quadrature with an error of 0.05 mag to represent systematic uncertainty, and these are included in the rest-frame photometry. Phases are given with respect to explosion.}
\begin{tabular}{|c|c|c|c|c|c|c|c|c|c|c|c|}
\hline
MJD & UT date & Rest-frame & g & (rest) & r & (rest) & i & (rest) & z & (rest)\\
& & phase (d) & (mag) & (mag) & (mag) & (mag) & (mag) & (mag) & (mag) & (mag) \\
\hline
56923.3 & 20140923 & -4.2 & >23.7 & -- & >23.5 & -- & -- & -- & --\\
56924.2 & 20140924 & -3.3 & >23.9 & -- & -- & -- & >23.9 & -- & >23.7 & -- \\
56931.1 & 20141001 & 3.3   & 18.98$\pm$0.005  & 18.86$\pm$0.06 & 19.13$\pm$0.01   & 18.99$\pm$0.06 & 19.3$\pm$0.01    & 19.14$\pm$0.05 & 19.44$\pm$0.01   & 19.24$\pm$0.05 \\
56934.4 & 20141004 & 6.4   & 18.115$\pm$0.003 & 17.89$\pm$0.05 & 18.281$\pm$0.004 & 18.06$\pm$0.05 & 18.42$\pm$0.01   & 18.2$\pm$0.05  & 18.6$\pm$0.01    & 18.35$\pm$0.05 \\
56936.3 & 20141006 & 8.3   & 17.504$\pm$0.002 & 17.51$\pm$0.05 & 17.727$\pm$0.002 & 17.68$\pm$0.05 & 17.911$\pm$0.002 & 17.86$\pm$0.05 & 18.077$\pm$0.003 & 18.02$\pm$0.05 \\
56943.2 & 20141013 & 14.9  & 17.137$\pm$0.001 & 17.04$\pm$0.05 & 17.278$\pm$0.001 & 17.16$\pm$0.05 & --            & 17.33$\pm$0.05 & 17.569$\pm$0.002 & 17.39$\pm$0.05 \\
56949.1 & 20141019 & 20.5  & 17.24$\pm$0.002  & 17.12$\pm$0.05 & 17.291$\pm$0.002 & 17.16$\pm$0.05 & 17.4$\pm$0.01    & 17.28$\pm$0.05 & 17.524$\pm$0.003 & 17.4$\pm$0.05  \\
56956.2 & 20141026 & 27.3  & 17.655$\pm$0.002 & 17.55$\pm$0.05 & 17.604$\pm$0.002 & 17.46$\pm$0.05 & 17.688$\pm$0.002 & 17.53$\pm$0.05 & 17.782$\pm$0.002 & 17.63$\pm$0.05 \\
56960.2 & 20141030 & 31.1  & 17.986$\pm$0.002 & 17.91$\pm$0.06 & 17.826$\pm$0.002 & 17.7$\pm$0.05  & 17.906$\pm$0.002 & 17.77$\pm$0.05 & --            & 17.89$\pm$0.05 \\
56973.0 & 20141112 & 43.4  & 19.405$\pm$0.005 & 19.43$\pm$0.07 & 18.726$\pm$0.003 & 18.64$\pm$0.06 & 18.716$\pm$0.003 & 18.58$\pm$0.05 & 18.78$\pm$0.004  & 18.68$\pm$0.05 \\
56980.0 & 20141119 & 50.1  & 20.55$\pm$0.01   & 20.63$\pm$0.07 & 19.57$\pm$0.01   & 19.51$\pm$0.07 & 19.49$\pm$0.01   & 19.37$\pm$0.06 & 19.43$\pm$0.01   & 19.38$\pm$0.05 \\
56987.0 & 20141126 & 56.8  & 21.53$\pm$0.02   & 21.67$\pm$0.07 & 20.32$\pm$0.01   & 20.25$\pm$0.06 & 20.28$\pm$0.01   & 20.16$\pm$0.06 & 20.08$\pm$0.01   & 20.1$\pm$0.05  \\
56990.1 & 20141129 & 59.7  & 21.85$\pm$0.14   & 21.95$\pm$0.19 & 20.51$\pm$0.06   & 20.51$\pm$0.08 & 20.51$\pm$0.17   & 20.49$\pm$0.08 & 20.37$\pm$0.04   & 20.35$\pm$0.06 \\
56991.1 & 20141130 & 60.7  & --            & 21.96$\pm$0.09 & 20.81$\pm$0.04   & 20.6$\pm$0.06  & 20.73$\pm$0.04   & 20.56$\pm$0.05 & 20.41$\pm$0.04   & 20.43$\pm$0.05 \\
56992.1 & 20141201 & 61.6  & 21.83$\pm$0.06   & --          & 20.65$\pm$0.02   & --          & 20.65$\pm$0.02   & --          & 20.36$\pm$0.02   & --          \\
57001.3 & 20141210 & 70.4  & 22.46$\pm$0.15   & 22.63$\pm$0.21 & 21.07$\pm$0.03   & 20.98$\pm$0.08 & 21.13$\pm$0.03   & 21.05$\pm$0.06 & 20.72$\pm$0.03   & 20.82$\pm$0.05 \\
57005.0 & 20141214 & 74.0  & 22.55$\pm$0.08   & 22.73$\pm$0.13 & 21.26$\pm$0.03   & 21.15$\pm$0.06 & --            & 21.36$\pm$0.07 & --            & 21.08$\pm$0.05 \\
57005.1 & 20141214 & 74.1  & --            & 23.36$\pm$0.25 & --            & 21.53$\pm$0.09 & 21.44$\pm$0.05   & 21.68$\pm$0.1  & 20.93$\pm$0.03   & 21.4$\pm$0.07  \\
57012.0 & 20141221 & 80.7  & 23.14$\pm$0.2    & --          & 21.63$\pm$0.05   & --          & 21.77$\pm$0.07   & --          & 21.24$\pm$0.05   & --          \\
57014.0 & 20141223 & 82.6  & 22.67$\pm$0.07   & 22.76$\pm$0.11 & 21.71$\pm$0.03   & 22.76$\pm$0.11 & 21.79$\pm$0.04   & 22.76$\pm$0.11 & 21.29$\pm$0.03   & 22.76$\pm$0.11 \\
57019.1 & 20141228 & 87.5  & 22.93$\pm$0.16   & 23.03$\pm$0.21 & 21.84$\pm$0.06   & 23.03$\pm$0.21 & 22.07$\pm$0.07   & 23.03$\pm$0.21 & 21.54$\pm$0.07   & 23.03$\pm$0.21 \\
57026.1 & 20150104 & 94.2  & 23.12$\pm$0.32   & 23.2$\pm$0.39  & 22.21$\pm$0.11   & 23.2$\pm$0.39  & 22.36$\pm$0.11   & 23.2$\pm$0.39  & 21.9$\pm$0.07    & 23.2$\pm$0.39  \\
57033.0 & 20150111 & 100.8 & 23.8$\pm$0.29    & 24.02$\pm$0.38 & 22.4$\pm$0.07    & 24.02$\pm$0.38 & 22.76$\pm$0.12   & 24.02$\pm$0.38 & 22.17$\pm$0.08   & 24.02$\pm$0.38 \\
57040.0 & 20150118 & 107.5 & 23.42$\pm$0.21   & 23.36$\pm$0.24 & 22.76$\pm$0.09   & 23.36$\pm$0.24 & 22.9$\pm$0.12    & 23.36$\pm$0.24 & 22.34$\pm$0.09   & 23.36$\pm$0.24 \\
57045.1 & 20150123 & 112.3 & 24.22$\pm$0.39   & 24.36$\pm$0.43 & 22.89$\pm$0.13   & 24.36$\pm$0.43 & 23.19$\pm$0.26   & 24.36$\pm$0.43 & 22.92$\pm$0.26   & 24.36$\pm$0.43 \\
57052.1 & 20150130 & 119.0 & 23.77$\pm$0.46   & 23.68$\pm$0.50  & 23.58$\pm$0.34   & 23.68$\pm$0.50  & 23.78$\pm$0.34   & 23.68$\pm$0.50  & 22.86$\pm$0.15   & 23.68$\pm$0.50  \\
\hline
\end{tabular}
\label{phot_table}
\end{table*}

Spectroscopy of DES14X2fna was taken between +17.5 and +52\,d (all phases stated in this paper are in the rest-frame and with respect to explosion epoch). These spectra were obtained with three different instruments: the AAOmega spectrograph at the Anglo-Australian Telescope (AAT) as part of the OzDES spectroscopic follow-up program, the Kast Double Spectrograph at Lick Observatory (LO) and the Blue Channel Spectrograph at the MMT Observatory. Details of the spectroscopic observations are in Table~\ref{spec_props}. Spectroscopic reductions were performed using standard procedures; the AAT spectrum was reduced following the procedure outlined in \citet{Childress2017}.

\begin{figure}
\centering
\includegraphics[width = \columnwidth]{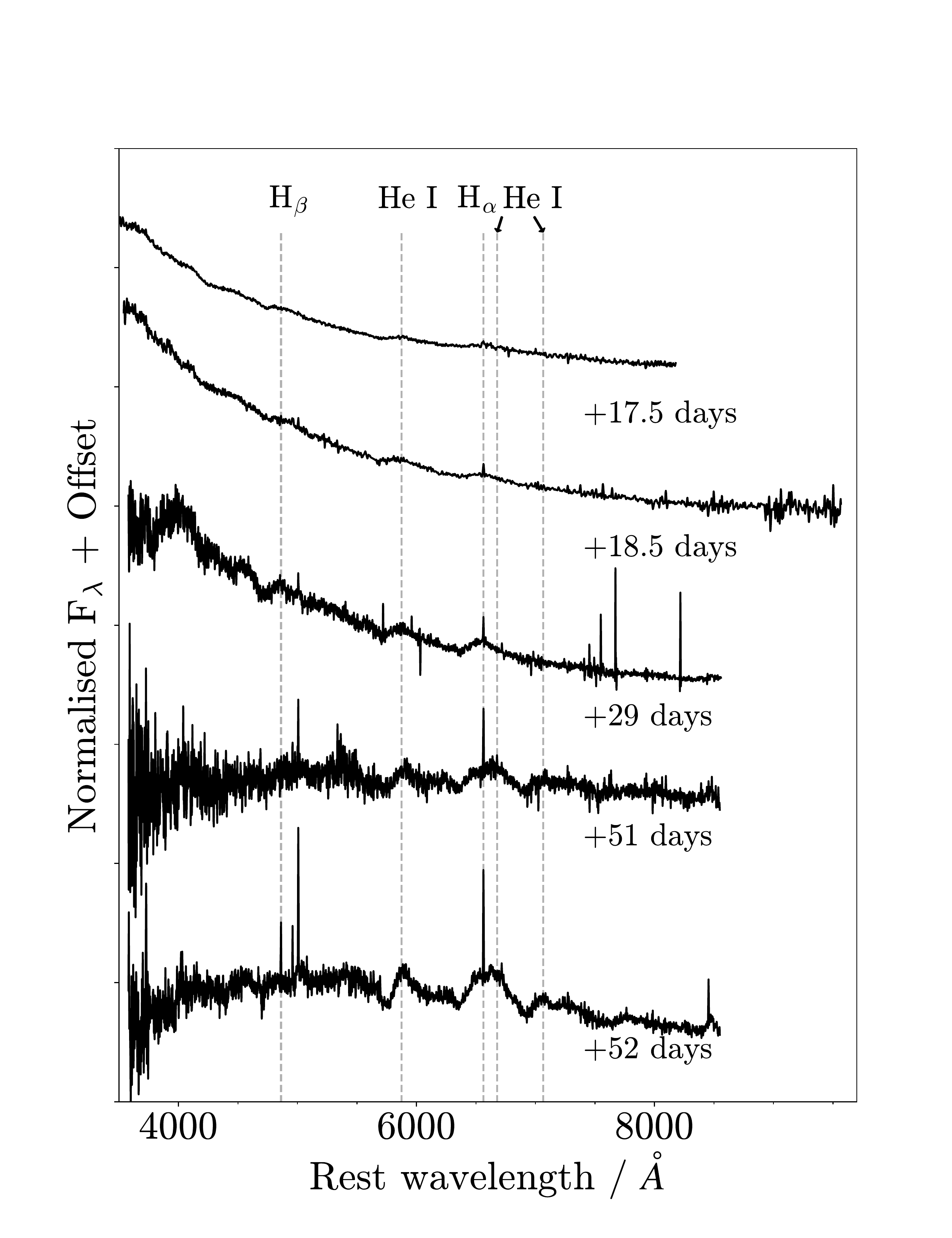}
\caption[]{Optical spectra of DES14X2fna. Each spectrum has been smoothed using a Savitzky-Golay filter. Spectra have been offset by an arbitrary amount for clarity, and corrected for redshift and Milky Way reddening using the extinction model of \citet{extinction}}
\label{spectra}
\end{figure}

The spectral evolution of DES14X2fna is shown in Fig.~\ref{spectra}. Note that as these spectra are only used for classification and calculating line velocities, they have not been calibrated to match photometry. Based on the presence of hydrogen at 18.5 days, it was initially classified as a SN II \citep{atel6591}. The appearance of helium at 52 days led to a reclassification as a SN IIb \citep{atel6789}.

\begin{table*}
\caption{Details of spectroscopy available for DES14X2fna. Phases are given with respect to explosion.}
\begin{tabular}{ |c|c|c|c|c| }
	\hline
	UT date & MJD & Rest-frame phase & Telescope & Range  \\
	&    & (d) & + Instrument & (\AA) \\
	\hline
	20141016 & 56946 & +17.5 & MMT+BCS & 3340-8550 \\
	20141017 & 56947 & +18.5 & LICK+Kast & 3400-10000 \\
	20141028 & 56958 & +29 & AAT+2dF/AAOmega & 3740-8950 \\
	20141120 & 56981 & +51 & AAT+2dF/AAOmega & 3740-8940 \\
	20141121 & 56982 & +52 & AAT+2dF/AAOmega & 3740-8950 \\
	\hline
\end{tabular}
\label{spec_props}
\begin{list}{}{}
\item NOTES:\\
BCS -- Blue Channel Spectrograph on MMT 6.5m telescope\\
Kast -- Kast Double Spectrograph on the 3m Shane telescope at Lick Observatory\\
2dF/AAOmega -- 2dF fibre positioner and AAOmega spectrograph on the 3.9-metre Anglo-Australian Telescope (AAT)\\
\end{list}
\end{table*}

\section{Characterising DES14X2fna}
\label{analyse}

\subsection{Host Galaxy}

\begin{figure}
\centering
\includegraphics[width = 0.9\columnwidth]{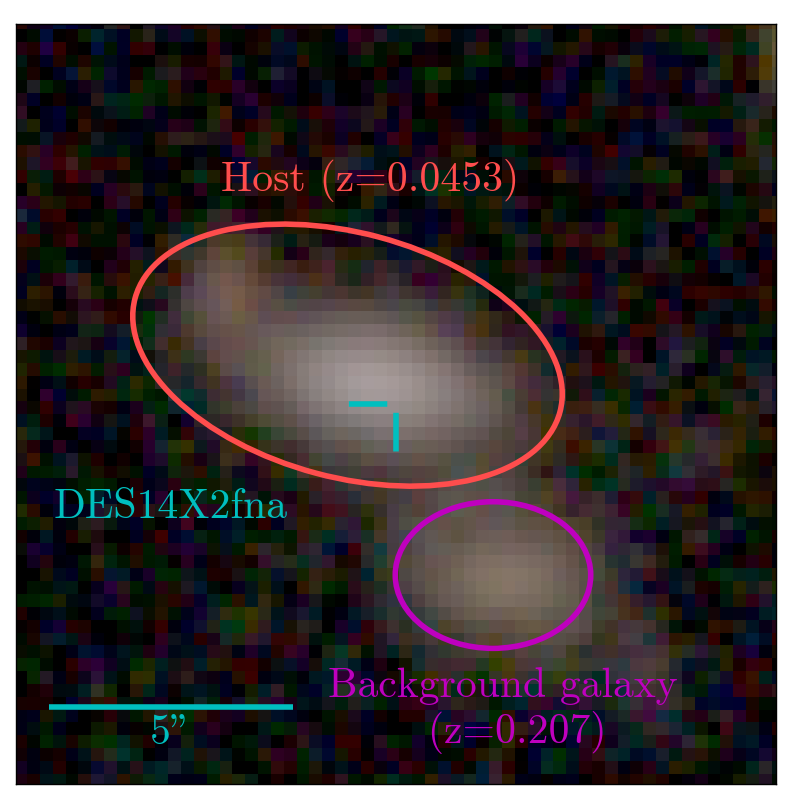}
\caption{A composite $gri$-band image of the host galaxy of DES14X2fna, from the stacked templates of the DES-SN field of \citet{stacks}. The location of the SN is indicated by the blue markers. The adjacent galaxy, shown by the purple circle, is at higher redshift and not in proximity to the host.}
\label{host_image}
\end{figure}

DES14X2fna was located in an anonymous host galaxy at a redshift of 0.0453. Assuming peculiar velocity dispersion of 200 km s$^{-1}$, this corresponds in our assumed cosmology to a distance of 200.7$\pm$3.0\,Mpc or a distance modulus of $\mu =$ 36.51$\pm$ 0.03\,mag. To infer global properties of the host, we use $griz$-band photometry from the deep stacked templates of the DES-SN fields described in \citet{stacks}. A composite $gri$-band image of the host galaxy from these templates is shown in Fig.~\ref{host_image}. The neighbouring galaxy is at $z=0.207$. This photometry corresponds to an absolute magnitude of $M_r=-16.35\pm0.03$. To estimate the stellar mass (\mstellar) and star formation rate (SFR), we fit stellar population synthesis models based on the templates of \citet{Bruzual2003}  with a \citet{Chabrier2003} initial mass function (IMF), as per \citet{Wiseman2020b}. We measure $\log_{10}$($M_*$/$M_\odot$)=$8.13^{+0.16}_{-0.07}$ and $\log_{10}$(SFR/$M_\odot$yr$^{-1}$)=$-1.53 ^ {\scriptscriptstyle +0.23} _{\scriptscriptstyle -0.47}$. We find consistent results when fitting with PÉGASE.2 templates and a Kroupa IMF using the method of \citet{Smith2020}.

We also derive abundance measurements from nebular emission lines in the host galaxy spectrum\footnote{Host spectrum of the host of DES14X2fna was taken in September 2018, by which time the SN had completely faded from view.} from the OzDES survey \citep{Lidman2020}, a spectroscopic redshift follow-up programme for DES. Emission line measurements and abundance calculations are performed using the method outlined in \citet{Wiseman2020b}. Briefly, we subtract the stellar continuum as well as any Balmer line absorption using the Penalized PiXel-Fitting software \citep[\textsc{pPXF};][]{ Cappellari2004,Cappellari2012,Cappellari2017} with the MILES library of single stellar populations \citep{Vazdekis2010}, and fit the resulting gas spectrum with Gaussian profiles. We use the line fluxes to derive metallicities based on a number of different calibrations: S2N2 \citep{Dopita2016}, N2 and O3N2 \citep{Pettini2004}, and R23 \citep{Kewley2004} (Table~\ref{host_props}). 

\begin{table}
\begin{center}
\caption[]{Properties of the host galaxy of DES14X2fna.}\begin{tabular}{ |c|c| }
	\hline
	Property & Host Value \\
	\hline
	M$_g$ & -16.01 $\pm$ 0.04 mag\\
	M$_r$ & -16.34 $\pm$ 0.04 mag\\
	M$_i$ & -16.50 $\pm$ 0.05 mag\\
	M$_z$ & -16.56 $\pm$ 0.05 mag\\
	Redshift & 0.0453 $\pm$ 0.0005 \\
	E(B-V)$_\text{MW}$ & 0.0225 mag\\
    log$_{10}$($M_*$/$M_\odot$) &   $8.13 _{\scriptscriptstyle -0.07} ^{\scriptscriptstyle +0.16}$ \\
    log$_{10}$(SFR/$M_\odot$yr$^{-1}$) &  $-1.53 _{\scriptscriptstyle -0.48} ^{\scriptscriptstyle +0.26}$ \\
    log$_{10}$(sSFR/yr$^{-1}$) &  $-9.66 _{\scriptscriptstyle -0.41} ^{\scriptscriptstyle +0.10}$ \\
    Metallicity: S2N2 D16       &   $7.76 _{\scriptscriptstyle -0.54} ^{\scriptscriptstyle +0.31}$ dex \\
    Metallicity: PP04 N2   &   $8.17 _{\scriptscriptstyle -0.17} ^{\scriptscriptstyle +0.11}$ dex \\
    Metallicity: PP04 O3N2    &   $8.19 _{\scriptscriptstyle -0.14} ^{\scriptscriptstyle +0.09}$ dex \\
    Metallicity: KK04 R23    &   $8.17 _{\scriptscriptstyle -0.23} ^{\scriptscriptstyle +0.20}$ dex \\
    \hline
\end{tabular}
\label{host_props}
\end{center}
\end{table}

Comparing our calculated mass of the host of DES14X2fna with previous studies of stellar masses of core-collapse SN hosts (e.g. Fig. 4 of \citealt{Galbany2018}, Fig. 11 of \citealt{stacks}) indicates that this is a low mass host. The PISCO sample of host galaxies presented in \citet{Galbany2018} contains 13 SNe IIb host environments - the host of DES14X2fna is over an order of magnitude less massive than any of these. This sample varies in PP04 O3N2 metallicity between approximately 8.4 to 8.7 and in log$_{10}$(sSFR/yr$^{-1}$) between approximately -11.2 to -9.5. DES14X2fna is lower metallicity than any of these hosts but falls in the upper end of the distribution in terms of sSFR.


In summary, the host of DES14X2fna is a low-mass, low-metallicity but relatively highly star-forming galaxy. The majority of core-collapse SNe are observed across a wide range of star-forming hosts \citep{general_host_metallicities}. However, the most energetic form of stripped envelope SN, SNe Ic-BL, are observed preferentially in low-mass, low-metallicity but highly star-forming environments similar to DES14X2fna \citep{IcBLhosts1,IcBLhosts2}.

\subsection{Photometry}
\label{fna_phot}

The rest-frame $griz$ light curves of DES14X2fna are shown in Fig.~\ref{LC}. After explosion, DES14X2fna rises to a peak $g$-band absolute magnitude of $\simeq-19.5$, and $\simeq-19.3$, $\simeq-19.2$ and $\simeq-19.1$ in $riz$-bands respectively. After peak $g$-band declines by $\sim$5 mags in $\sim$40 days, while $riz$ decline by $\sim$3 mags over the same period. After approximately 57 days, each band appears to show a roughly linear decline. GP-interpolation can be used to estimate the rise time and peak absolute magnitude, but does not directly provide us with uncertainties on rise time. We estimate these using a Monte Carlo approach, randomising the rest-frame photometry within error bars and GP-interpolating this to estimate the rise time and peak absolute magnitude. This is repeated 1000 times, with the mean and standard deviation taken as the final values and uncertainties. These values are shown in Table~\ref{peak_phot_table}.
\begin{table}
\begin{center}
\caption{Peak absolute magnitudes of DES14X2fna estimated from Gaussian Process interpolation. Statistical errors in rise time estimated from MC approach, systematic errors correspond to uncertainty in explosion epoch between last non-detection and first detection in rest-frame.}
\begin{tabular}{ |c|c|c| }
	\hline
	Filter & Peak absolute & Rise time\\
	&magnitude (mag)&(d)\\
	\hline
	$g$ & -19.47 $\pm$ 0.06 & 16.67 $\pm$ 0.53$_\text{stat}$ $\pm$ 3.35$_\text{sys}$\\
	$r$ & -19.37 $\pm$ 0.05 & 18.00 $\pm$ 0.56$_\text{stat}$ $\pm$ 3.35$_\text{sys}$\\
	$i$ & -19.23 $\pm$ 0.06 & 18.90 $\pm$ 0.58$_\text{stat}$ $\pm$ 3.35$_\text{sys}$\\
	$z$ & -19.14 $\pm$ 0.05 & 18.35 $\pm$ 0.57$_\text{stat}$ $\pm$ 3.35$_\text{sys}$ \\
    \hline
\end{tabular}
\label{peak_phot_table}
\end{center}
\end{table}

To characterise the light curve of DES14X2fna relative to the population of SNe IIb, we have gathered a comparison sample of spectroscopically confirmed SNe IIb. We have selected objects with publicly-available photometry with good coverage around peak and a well-constrained explosion epoch, either through light curve modelling or a short period between last non-detection and first detection (15 days in the maximum in our sample). This leaves us with a sample of 22 SNe IIb (see Table~\ref{IIb_table}).

\begin{figure*}
\centering
\includegraphics[width = \textwidth]{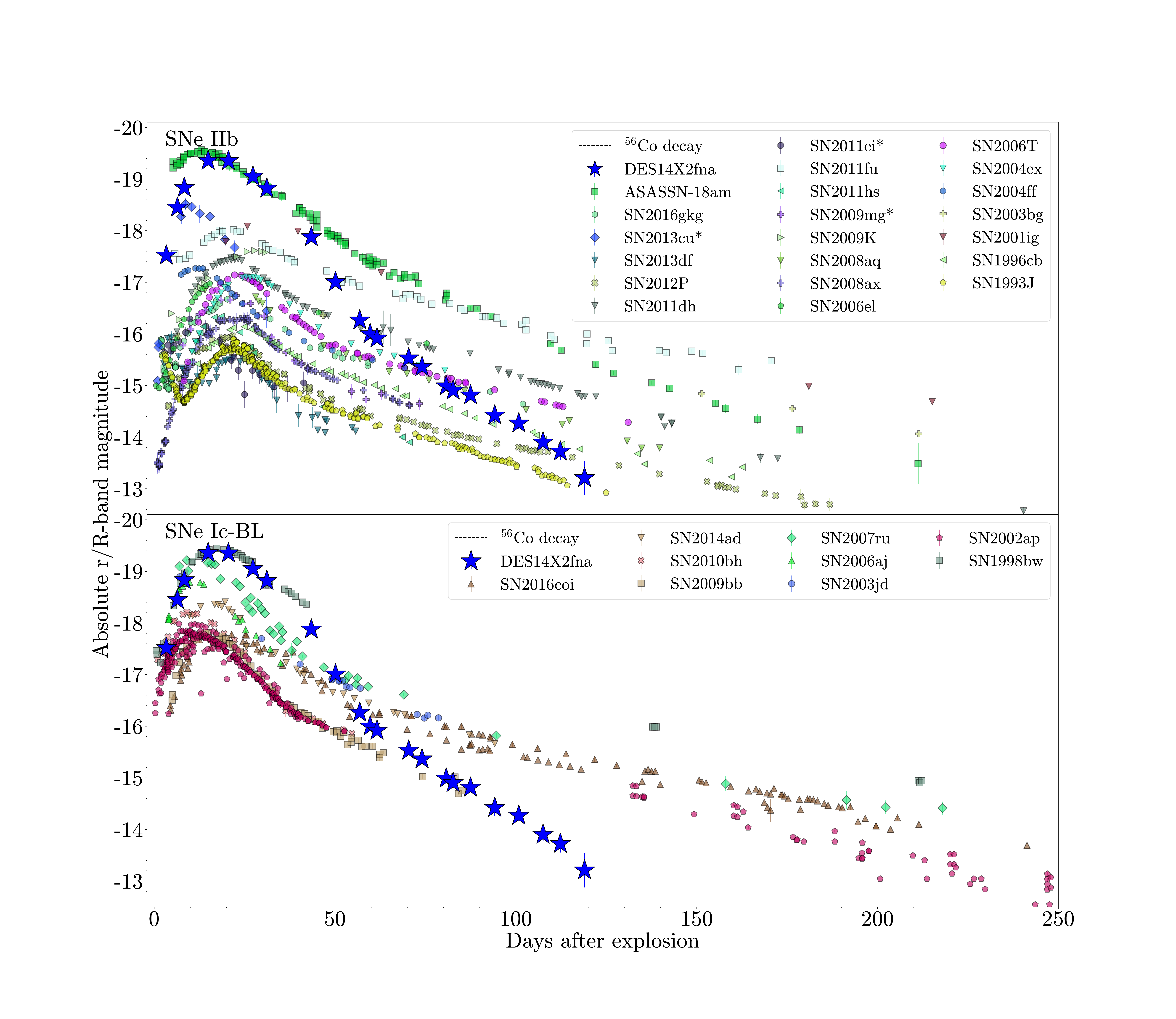}
\caption{\textbf{Upper: } $r$-band light curve of DES14X2fna alongside the $r$/$R$-band light curves of our SN IIb comparison sample. The dashed line indicates the decline rate expected for fully-trapped $^{56}$Co decay. Objects denoted with * in the legend show $V$-band photometry due to a lack of available $r$/$R$-band data \textbf{Lower: } Same as upper panel but for a comparison sample of SNe Ic-BL. All light curves are corrected for MW extinction, and DES14X2fna is k-corrected to the rest-frame.}
\label{comp_LC}
\end{figure*}

The top panel of Fig.~\ref{comp_LC} shows the rest-frame $r$-band light curve of DES14X2fna along with $r$/$R$-band light curves of our SNe comparison sample. Note that SN~2009mg, SN~2011ei and SN~2013cu lack sufficient $r$/$R$-band coverage and so we have used $V$-band photometry as the closest wavelength band available to $R$. For each object we use GP-interpolation to estimate the rise time and peak absolute magnitude. We  estimate the epoch at which the linear decline phase of the light curve begins for each SN, $t_\text{tail}$ and the absolute magnitude at this epoch, $M_\text{tail}$, by eye, where these could be inferred from the data. In addition, we perform linear fits to calculate the tail decline rate where possible. Details and parameter values of each SN in our comparison sample are given in Table \ref{IIb_table}. We have corrected the light curves for redshift and Milky Way extinction based on values reported in the respective papers outlined in this table. However, as these objects are all at low redshift, we do not K-correct these light curves to the rest frame since the effect will be small. Uncertainties in quoted peak absolute magnitude ($M_\text{peak,MW}$) incorporate uncertainty in GP-interpolation to observed photometry and Milky Way extinction correction. Distances are estimated by cross-matching SNe with their hosts using the Nasa Extragalactic Database (NED\footnote{\url{http://ned.ipac.caltech.edu/}}) and obtaining a luminosity distance in our assumed cosmology corrected for Virgo infall. We choose this approach for consistency as some of these hosts do not have redshift-independent distance estimates and where they do these are often calculated for different cosmologies. The exception to this is SN~1993J, as the negative redshift of the host means that a redshift distance is not appropriate - instead, we adopt the literature value of 2.9$\pm$0.4 Mpc from \citet{93J_dist}. Distance uncertainties are included in this analysis. Where data is available, we correct for host extinction and factor this into the peak absolute magnitude $M_\text{peak,host}$. However, as this data was not available for all objects and only an upper limit in some cases we do not correct for host extinction when considering bolometric luminosities in order to be consistent across all objects.

The most noticeable feature of DES14X2fna is its very high peak luminosity, ~0.2 mag fainter than the very luminous ASASSN-18am at peak but nearly one mag brighter than the  next brightest object in the sample, SN~2013cu. It is important to note that host extinction can make a significant difference to the luminosity of each object, as in the case of SN2008ax which shows an increase in peak luminosity of greater than 1 mag. However, as most objects have a host extinction considerably less than this and DES14X2fna is significantly more luminous than most of the sample, DES14X2fna still stands out for its considerable luminosity even when accounting for this possibility. DES14X2fna rises to peak in $r$-band in 18 days, similar to ASASSN-18am and SN~2011hs and fairly typical for SNe IIb based on the sample of SN IIb rise times presented in Fig. 4 of \citet{Pessi19}. After rising to maximum, DES14X2fna declines at a similar rate to ASASSN-18am until ~30 days after explosion, after which DES14X2fna begins to decline far more rapidly. 

DES14X2fna has a single-peaked light curve similar to SN~2008ax and unlike the well-studied SN~1993J. However, as there is a period of $\simeq$7 days between the last non-detection and first detection it is possible that an initial peak did occur but was not observed, as this has been observed to last only a few days in previous SNe \citep{IIb_sample13, IIb_sample26,Tartaglia2017}.


\begin{table*}
\scriptsize
\begin{center}
\caption{Details and light curve properties of our SN IIb comparison sample. These properties were calculated using $r$/$R$-band photometry with the exception of SNe marked with *, for which $V$-band light curves were used due to a lack of $r$/$R$-band data. Missing host extinction values have been labelled as either negligible based on spectroscopy or N/A if no host extinction data was available. All presented magnitudes are in AB system. Where SN explosion dates were calculated using the dates of first detection and last non-detection, rise time uncertainties have quoted systematic errors as in table \ref{peak_phot_table}. These values are not quoted where explosion epochs were sourced from literature using other methods.}
\scriptsize
\setlength{\tabcolsep}{5pt}
\begin{tabular}{|l|c|c|c|c|c|c|c|c|c|c|c|l|}
\hline
SN & Redshift & E(B-V)$_\text{MW}$ & E(B-V)$_\text{host}$ & Rise time & M$_\text{peak,MW}$ & M$_\text{peak,host}$ & t$_\text{tail}$ & M$_\text{tail}$ & Tail decline rate & References \\
 & & (mag) & (mag) & (d) & (mag) & (mag) & (d) & (mag) & (mag 100d$^{-1}$) & \\
\hline
DES14X2fna & 0.0453    & 0.0225$\pm$0.0003 & Negligible & 18.00$\pm$0.56$_\text{stat}\pm$3.35$_\text{sys}$ & -19.37$\pm$0.05 & -- & 60 & -16 & 4.38$\pm$0.10 & --    \\
ASASSN-18am & 0.0301 & 0.0086$\pm$0.0011 & Negligible & 18.89$\pm$0.42$_\text{stat}\pm0.39_\text{sys}$ & -19.53$\pm$0.04 & -- & 60 & -17.3 & 3.03$\pm$0.06 & (1) \\
SN~2016gkg  & 0.0049 & 0.0166$\pm$0.0002 & 0.09$^{+0.08}_{-0.07}$ & 20.45$\pm1.07_\text{stat}\pm0.16_\text{sys}$ & -16.69$\pm$0.16 & -16.97$^{+0.30}_{-0.27}$ & 43 & -16 & 1.63$\pm$0.08 & (2), (3), (4), (5), (6), (7) \\
SN~2013cu*   & 0.0252  & 0.0105$\pm$0.0003 & N/A & 9.83$\pm$1.07$_\text{stat}\pm$0.97$_\text{sys}$ & -18.51$\pm$0.17 & -- & -- & -- & -- & (2), (8)   \\
SN~2013df   & 0.0024  & 0.0168$\pm$0.0002 & 0.081$\pm$0.016 & 22.28$\pm$0.67$_\text{stat}$ & -15.53$\pm$0.16 & -15.78$\pm$0.17 & 40  & -14.8 & 1.96$\pm$0.02 & (2), (9), (10), (11), (12) \\
SN~2012P    & 0.0045  & 0.0437$\pm$0.0005 & 0.29$^{+0.08}_{-0.05}$ & 20.33$\pm$0.19$_\text{stat}$ & -15.93$\pm$0.16 & -- & [50,60] & [-14,-14.2] & 1.44$\pm$0.02 & (2), (13)  \\
SN~2011dh & 0.0016 & 0.0309$\pm$0.0017 & <0.05 & 21.90$\pm0.20_\text{stat}\pm0.88_\text{sys}$ & -17.47$\pm$0.15 & -- & 48 & -15.8 & 2.02$\pm$0.03 & (2), (9), (14), (15), (16) \\
SN~2011ei*   & 0.0093  & 0.0505$\pm$0.0008 & 0.18 & 11.87$\pm$0.71$_\text{stat}$ & -16.14$\pm$0.17 & -16.70$\pm$0.17  & -- & -- & (2), (17)  \\
SN~2011fu   & 0.0185  & 0.0648$\pm$0.0008 & 0.15$\pm$0.11 & 20.49$\pm$0.54$_\text{stat}$ & -18.02$\pm$0.15 & -18.49$\pm$0.38 & [40,45] & [-17.2,-17.4] & 1.69$\pm$0.05 & (18), (19)   \\
SN~2011hs   & 0.0057 & 0.0104$\pm$0.0004 & 0.16$\pm$0.07 & 17.89$\pm$0.31$_\text{stat}$ & -16.35$\pm$0.15 & -16.85$\pm$0.26 & 35 & -15.1 & 2.14$\pm$0.02 & (2), (20)  \\
SN~2009K & 0.0117 & 0.0491$\pm$0.0014 & N/A & 27.44$\pm0.39_\text{stat}\pm1.21_\text{sys}$ & -17.61$\pm$0.15 & -- & --  & -- & -- & (21), (22), (23), (24) \\
SN~2009mg*   & 0.0076 & 0.0388$\pm$0.0005 & 0.09$\pm$0.02 & 21.07$\pm$0.95$_\text{stat}$ & -16.69$\pm$0.16 & -16.97$\pm$0.17 & -- & -- & -- & (2), (25)  \\
SN~2008aq   & 0.008 & 0.0386$\pm$0.0009 & 0.027 & 20.11$\pm$0.47$_\text{stat}$ & -16.91$\pm$0.15 & -17.00$\pm$0.15 & 47 & -15.7 & 2.18$\pm$0.04 & (2), (21), (22), (26) \\
SN~2008ax & 0.0019 & 0.0188$\pm$0.0002 & 0.38$\pm$0.1 & 23.89$\pm$1.03$_\text{stat}$ & -16.32$\pm$0.15 & -17.50$\pm$0.34 & 40 & -15.3 & 2.13$\pm$0.03 & (2), (22), (27), (28) \\
SN~2006el   & 0.017 & 0.0975$\pm$0.0012 & N/A & 22.19$\pm$1.05$_\text{stat}$ & -17.134$\pm$0.17 & -- & --  & -- & -- & (22), (29)  \\
SN~2006T    & 0.0081 & 0.0643$\pm$0.0007 & N/A & 22.84$\pm0.22_\text{stat}\pm7.63_\text{sys}$ & -17.15$\pm$0.15 & -- & 48  & -15.9 & 1.79$\pm$0.02 & (2), (21), (22) \\
SN~2004ex   & 0.018 & 0.0184$\pm$0.0016 & N/A & 25.60$\pm0.55_\text{stat}\pm10.43_\text{sys}$ & -17.10$\pm$0.15 & -- & 54 & -15.9 & 2.06$\pm$0.08 & (22)  \\
SN~2004ff   & 0.023 & 0.0274$\pm$0.001 & N/A & 11.83$\pm$0.37$_\text{stat}$ & -17.27$\pm$0.15 & -- & 40  & -16 & 1.88$\pm$0.08 & (21), (29), (30) \\
SN~2003bg   & 0.0046 & 0.018$\pm$0.001 & Negligible & 49.45$\pm$0.77$_\text{stat}$ & -16.36$\pm$0.16 & -- & -- & -- & 1.29$\pm$0.06 & (31)  \\
SN~2001ig   & 0.0031 & 0.0089$\pm$0.0004 & <0.09 & 30.20$\pm$3.57$_\text{stat}$ & -18.01$\pm$0.16 & -- & --  & -- & -- & (32) \\
SN~1996cb   & 0.0024 & 0.0261$\pm$0.0005 & <0.12 & 24.06$\pm$0.66$_\text{stat}$ & -16.10$\pm$0.16 & -- & 44  & -15.6 & 2.21$\pm$0.03 & (33)  \\
SN~1993J    & -0.0001 & 0.069$\pm$0.0001 & 0.12$\pm$0.07 & 21.26$\pm$3.86$_\text{stat}$ & -16.85$\pm$0.30 & -17.22$\pm$0.37 & 40 & -15.8 & 2.18$\pm$0.02 & (9), (34), (35), (36),  \\
& & & & & & & & & & (37), (38), (39), (40), \\
& & & & & & & & & & (41), (42), (43) \\
\hline
\end{tabular}
\begin{list}{}{}
\item References: (1) \citet{asassn}; (2) \citet{IIb_sample2}; (3) \citet{IIb_sample30}; (4) \citet{IIb_sample31}; (5) \citet{IIb_sample32}; (6) \citet{IIb_sample33}; (7) \citet{IIb_sample34}; (8) \citet{IIb_sample1}; (9) \citet{IIb_sample5}; (10) \citet{IIb_sample17}; (11) \citet{IIb_sample18}; (12) \citet{IIb_sample19}; (13) \citet{IIb_sample29}; (14) \citet{IIb_sample24}; (15) \citet{IIb_sample25}; (16) \citet{IIb_sample26}; (17) \citet{IIb_sample42}; (18) \citet{IIb_sample3}; (19) \citet{IIb_sample4}; (20) \citet{IIb_sample27}; (21) \citet{IIb_sample20}; (22) \citet{IIb_sample21}; (23) \citet{IIb_sample40}; (24) \citet{IIb_sample41}; (25) \citet{IIb_sample36}; (26) \citet{IIb_sample35}; (27) \citet{IIb_sample22}; (28) \citet{IIb_sample23}; (29) \citet{IIb_sample38}; (30) \citet{IIb_sample39}; (31) \citet{IIb_sample16}; (32) \citet{IIb_sample43}; (33) \citet{IIb_sample28}; (34) \citet{IIb_sample6}; (35) \citet{IIb_sample7}; (36) \citet{IIb_sample8}; (37) \citet{IIb_sample9}; (38) \citet{IIb_sample10}; (39) \citet{IIb_sample11}; (40) \citet{IIb_sample12}; (41) \citet{IIb_sample13}; (42) \citet{IIb_sample14}; (43) \citet{IIb_sample15}.
\end{list}
\label{IIb_table}
\end{center}
\end{table*}

Motivated by the bright peak luminosity of DES14X2fna, we also compare to SNe Ic-BL to study any resemblance to this very energetic class of stripped envelope SNe. Following the method for our SN IIb sample, we form a similar comparison sample of SNe Ic-BL (Table~\ref{Ic_table}), and light curves shown in the lower panel of Fig.~\ref{comp_LC}. Overall, the peak of DES14X2fna resembles that of a SN Ic-BL more closely than a typical SN IIb. Most notably, in $r$-band DES14X2fna matches SN~1998bw well until $\simeq30$ days after explosion. After this time DES14X2fna declines more rapidly than SN~1998bw.

\begin{table}
\caption{Our SN Ic-BL comparison sample.}
\begin{tabular}{|c|c|c|l|}
\hline
SN & Redshift & $E(B-V)_\textrm{MW}$ & References \\
\hline
SN~2016coi & 0.003646 & 0.0737 & (1) \\
SN~2014ad  & 0.0057   & 0.038  & (2)  \\
SN~2010bh  & 0.0593   & 0.1004 & (3)  \\
SN~2009bb  & 0.0104   & 0.0847 & (4), (5) \\
SN~2007ru  & 0.0155   & 0.2217 & (6), (7) \\
SN~2006aj  & 0.033023 & 0.1261 & (6)  \\
SN~2003jd  & 0.019    & 0.0378 & (6)  \\
SN~2002ap  & 0.002108 & 0.0616 & (6), (8), (9), (10), (11) \\
SN~1998bw  & 0.0085   & 0.0494 & (12), (13), (14), (15) \\
\hline
\end{tabular}
(1) \citet{Ic_sample18}; (2) \citet{Ic_sample7}; (3) \citet{Ic_sample8}; (4) \citet{IIb_sample20}; (5) \citet{Ic_sample10}; (6) \citet{IIb_sample21}; (7) \citet{Ic_sample19}; (8) \citet{Ic_sample3}; (9) \citet{Ic_sample4}; (10) \citet{Ic_sample5}; (11) \citet{Ic_sample6}; (12) \citet{Ic_sample14}; (13) \citet{Ic_sample15}; (14) \citet{Ic_sample16}; (15) \citet{Ic_sample17}
\label{Ic_table}
\end{table}

\begin{figure*}
\centering
\includegraphics[width = \textwidth]{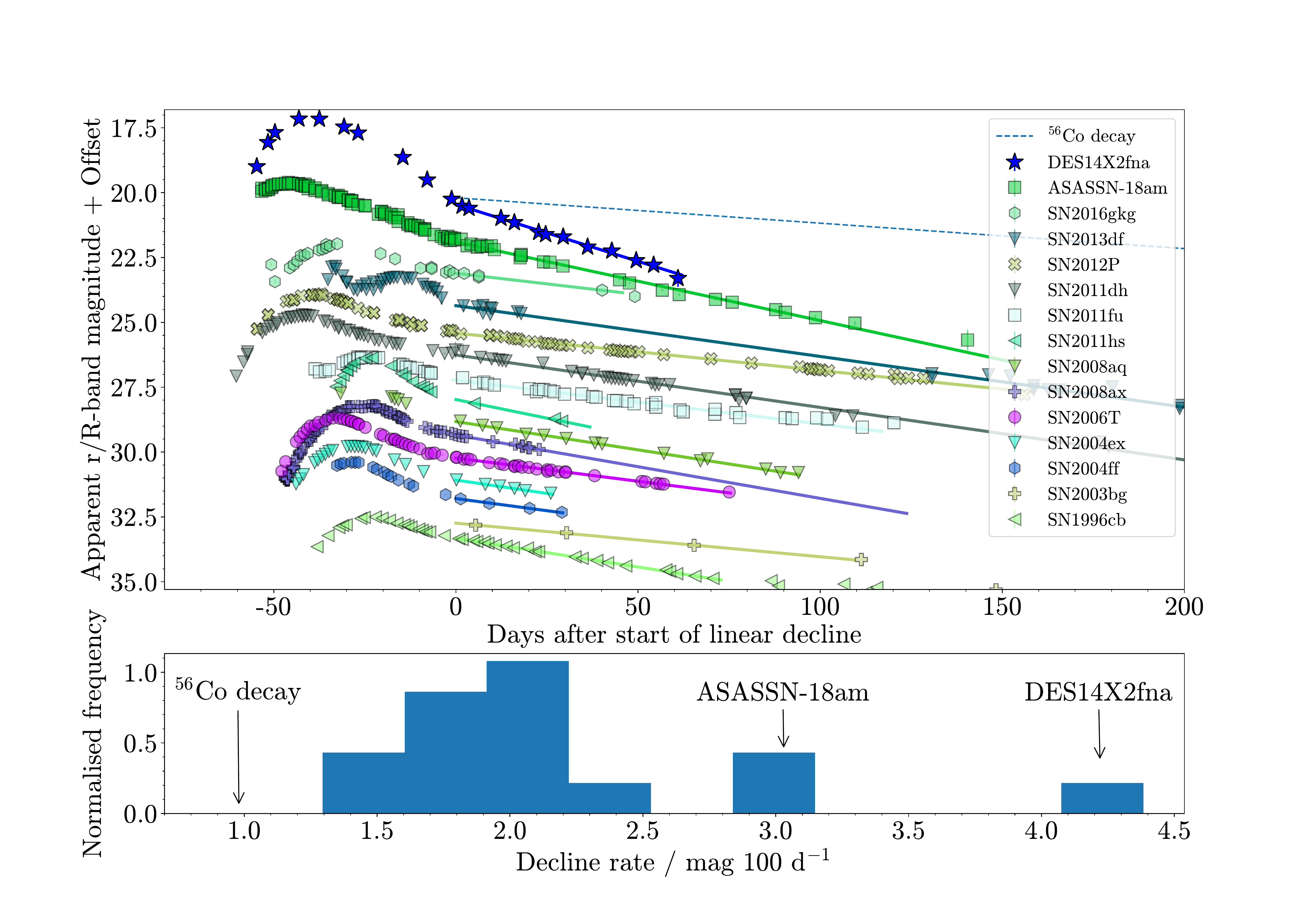}
\caption{\textbf{Upper:} $r$-band light curve of DES14X2fna and the $r$/$R$-band light curves of the SNe in the SN IIb comparison sample that exhibit a linear decline. The solid lines show a linear fit to the post-peak linear decline of each SN. \textbf{Lower:} Histogram showing $r$/$R$-band decline rates estimated from the linear fits in upper panel, with DES14X2fna annotated.}
\label{declines}
\end{figure*}

As is evident from both panels of Fig.~\ref{comp_LC}, after peak DES14X2fna exhibits a very rapid tail decline compared to both SNe IIb and SNe Ic-BL. The top panel of Fig.~\ref{declines} focuses on this decline phase for each SN in our SN IIb comparison sample where a tail is apparent. For each SN in the plot, we fit a line to the tail to estimate the decline rate (Table \ref{IIb_table}). DES14X2fna has an $r$-band decline rate of 4.38$\pm0.10$\,mag (100d)$^{-1}$, significantly faster than ASASSN-18am, the next fastest decliner in the sample ($3.03\pm0.06$\,mag (100d)$^{-1}$). The lower panel of Fig.~\ref{declines} also shows a histogram of the estimated decline rates for our sample, where DES14X2fna is a clear outlier.

It is clear that DES14X2fna exhibits both a very luminous peak and a faster tail evolution than other previously observed SNe IIb. If we are to assume that this object follows the canonical $^{56}$Ni decay model for a SN IIb, high $^{56}$Ni and ejecta masses would be required to power the high maximum luminosity and broad peak of DES14X2fna. However, a high ejecta mass would be expected to lead to complete trapping of the $\gamma$-rays produced in $^{56}$Co decay at late times as there would be significant surrounding material to absorb these $\gamma$-rays. Not only does DES14X2fna decline at a rate far faster than expected for fully-trapped $^{56}$Co decay, it declines nearly 1.5 mag (100d)$^{-1}$ faster than any other SN IIb in our comparison sample. This raises the question as to whether the light curve of DES14X2fna is consistent with a $^{56}$Ni decay model or if an alternative mechanism to power the light curve is required.

\subsection{Bolometric Luminosity}
\label{bol_lum}

We next consider the bolometric and pseudo-bolometric light curves of DES14X2fna. Observed $griz$ light curves are converted to luminosities by fitting with a black body curve to compute a spectral energy distribution (SED). The exact values and uncertainties in bolometric luminosities were estimated using a Monte Carlo approach:

\begin{itemize}
    \item An initial black body fit is carried out to estimate photometric radius and temperature along with their uncertainties.
    \item Randomised radius and temperature values are then drawn from a Gaussian distribution using the best-fit values and uncertainties and used to generate a randomised black body SED.
    \item Each randomised SED is integrated over the wavelength range covered by $griz$ bands to produce a pseudo-bolometric luminosity, $L_{griz}$. A bolometric luminosity $L_{\text{bol}}$ is estimated by integrating the fitted SED over all wavelengths.
    \item This process is then repeated, with the final values of $L_{griz}$ and $L_{\text{bol}}$ as well as their uncertainties calculated from the mean and standard deviation of values from all of the randomised black body SEDs.
\end{itemize}

We also considered estimating bolometric luminosity using our mangled SED models and the bolometric corrections outlined in \citet{Lyman14}, but find that these obtain consistent results with our black body approach. As a result, we use black body fits since these provide information on photospheric temperature and radius as well as luminosity. Of course, the black body approximation is not valid at later times as the SN enters the nebular phase hence we do not estimate bolometric luminosity more than 120 days after explosion, after our last observation of DES14X2fna. At later times the decreasing quality of a black body fit is reflected in larger fit uncertainties and hence much larger uncertainties in $L_{bol}$ and $L_{griz}$. In addition to this, further uncertainty in $L_{bol}$ arises from the lack of flux information beyond optical wavelengths, particularly in the UV. Without observations at these wavelengths, we are not able to quantify this effect but acknowledge that it will serve to increase our uncertainties.

\begin{figure*}
\centering
\includegraphics[width = \textwidth]{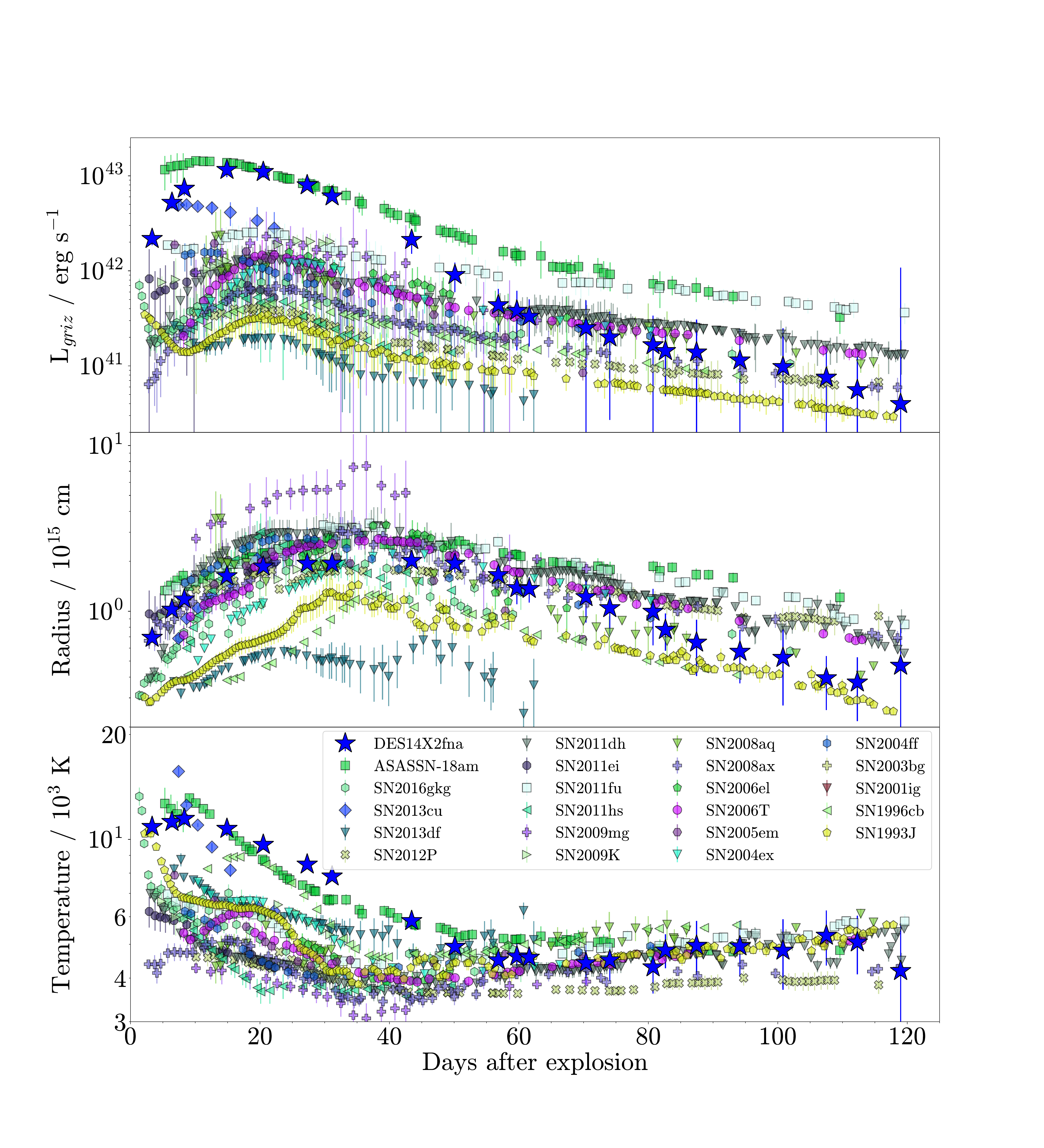}
\caption{\textbf{Upper:} Bolometric light curves of DES14X2fna and our SN IIb comparison sample, estimated from black body fits to observed photometry. \textbf{Middle:} Photospheric radii for DES14X2fna and our SN IIb comparison sample estimated from black body fits. \textbf{Lower:} Effective temperatures for DES14X2fna and our SN IIb comparison sample.}
\label{BL}
\end{figure*}

To calculate bolometric light curves for our SNe IIb comparison sample, we make cuts to select only SNe with observations in at least 3 photometric bands around peak luminosity to allow reasonable black body fits. Unlike DES14X2fna, our sample of  SNe IIb includes objects with photometric data at very different epochs and phases of the light curve, which restricts the epochs at which we can calculate bolometric luminosities. At each epoch with available photometry, other observed bands were GP-interpolated to provide simultaneous photometry, after which the same procedure outlined for DES14X2fna was carried out to estimate black body fit parameters and bolometric luminosities.
To ensure consistency with our estimates of the pseudo-bolometric luminosity of DES14X2fna, we use the same wavelengths limits in our integration regardless of which photometric bands were available for each SN. The bolometric luminosities, photospheric radii and temperatures obtained from our black body fits are shown in Fig.~\ref{BL}.

As with the $r$/$R$-band photometric data, aside from ASASSN-18am DES14X2fna is the most luminous object in the SN IIb sample and has a relatively broad light curve peak. We can estimate the true peak bolometric luminosity and rise time using GP-interpolation, which gives a peak luminosity of $L_{\text{bol}} = (2.44\pm0.41)\times10^{43}$ erg~s$^{-1}$ and $L_{griz} = (1.17\pm0.08)\times10^{43}$ erg~s$^{-1}$, with a rise time to peak, $t_p$ of 15.38 days. With the exception of ASASSN-18am and SN~2013cu, DES14X2fna is brighter at peak than any other SN in the sample by more than a factor of 2. This reiterates our findings from broad-band photometry: DES14X2fna is very luminous at maximum, comparable only to ASASSN-18am, and has a relatively broad light curve peak.

The photospheric radius of DES14X2fna shows an initial fast rise, reaching (1.86 $\pm$ 0.08) $\times\ 10^{15}$ cm after 20.5 days, suggesting a photospheric velocity of $\sim$7,900 km s$^{-1}$. The radius then remains roughly constant for $\sim30$ days, before slowly decreasing again. This radius evolution is fairly typical for a SN IIb, most closely resembling SN~2011dh and SN~2004ff. ASASSN-18am also shows a similar radius evolution to DES14X2fna at early times, although declines more slowly after peak.

DES14X2fna and ASASSN-18am both show a similar temperature evolution, rising to $\sim12,000$ K at peak after $\sim8-9$ days, before declining roughly linearly to $\sim5000$ K after $\sim$50 days and staying roughly constant thereafter, although ASASSN-18am declines more rapidly between $\sim10-20$ days. While SN~2016gkg and SN~2013cu do reach comparable temperatures, this occurs very soon after explosion due to shock cooling while DES14X2fna and ASASSN-18am show temperatures in excess of the rest of the SNe IIb sample until 50 days after explosion, with a far more prolonged temperature decline.

\subsection{Spectroscopy}

Spectroscopy of DES14X2fna is shown in Fig.~\ref{spectra}. After $\sim17.5$ days, weak H$\beta$ and \ion{He}{i} 5876 lines become visible. Approximately one day later a more noticeable broad H$\alpha$ feature is visible. From $\sim29$ days, the features become more prominent, with H$\alpha$, H$\beta$ and \ion{He}{i} 5876 all visible. These are still present in the later spectra at $\sim50$ days, albeit with H$\beta$ becoming increasingly noisy, and further \ion{He}{i} features at 6678\AA\ and 7065\AA\ also appear in these later spectra.

Where possible, we have estimated the expansion velocities of these lines from their P-Cygni profiles, fitting a Gaussian with a pseudo-continuum to estimate the minimum and using this to infer velocity. We follow the process outlined in \citet{Kate_continuum}, fitting to a small wavelength range around each feature and considering a 30 \AA\ range for the cutoff on either side of the feature when searching for the best fit. The velocity evolution of DES14X2fna is shown in Table \ref{exp_velocities}.

\begin{table}
\begin{center}
\begin{tabular}{ |c|c|c|c| }
	\hline
	Line & Phase & v$_{exp}$ / km s$^{-1}$  \\
	\hline
	H$_\alpha$ & +29 & 8670 $\pm$ 660 \\
	-- & 51 & 8880 $\pm$ 740 \\
	-- & 52 & 9540 $\pm$ 510 \\
	H$_\beta$ & +17.5 & 8250 $\pm$ 150 \\
	-- & 29 & 9110 $\pm$ 450 \\
	-- & 52 & 9620 $\pm$ 1830 \\
	\ion{He}{I} 5876 & +17.5 & 9250 $\pm$ 520 \\
	-- & 29 & 8480 $\pm$ 300 \\
	-- & 51 & 6550 $\pm$ 970 \\
	-- & 52 & 6220 $\pm$ 490 \\
	\ion{He}{I} 7065 & +51 & 6620 $\pm$ 770 \\
	-- & 52 & 6220 $\pm$ 490 \\
	\hline
\end{tabular}
\end{center}
\caption{Expansion velocities of DES14X2fna at different phases calculated from position of minimum of P-Cygni profile}
\label{exp_velocities}
\end{table}

Overall, H$\alpha$ and H$\beta$ maintain a relatively constant velocity throughout all of our spectra of roughly 9000\,km\,s$^{-1}$. \ion{He}{I} 5876 has a velocity of 9254 $\pm$ 517 km s$^{-1}$ after ~17.5 days, comparable to the H lines, before decreasing gradually in each spectrum and reaching 6074 $\pm$ 119 km s$^{-1}$ after ~52 days. \ion{He}{I} 7065 is not visible in the early spectra but has velocities of 6616 $\pm$ 766 km s$^{-1}$ and 6219 $\pm$ 489 in the final two spectra respectively, similar to \ion{He}{I} 5876. The decreasing velocity of He but not H could result if almost all H present is located in the outer ejecta, meaning that there is little or no H below 8000-9000\,km\,s$^{-1}$.

\begin{figure}
\centering
\includegraphics[width = \columnwidth]{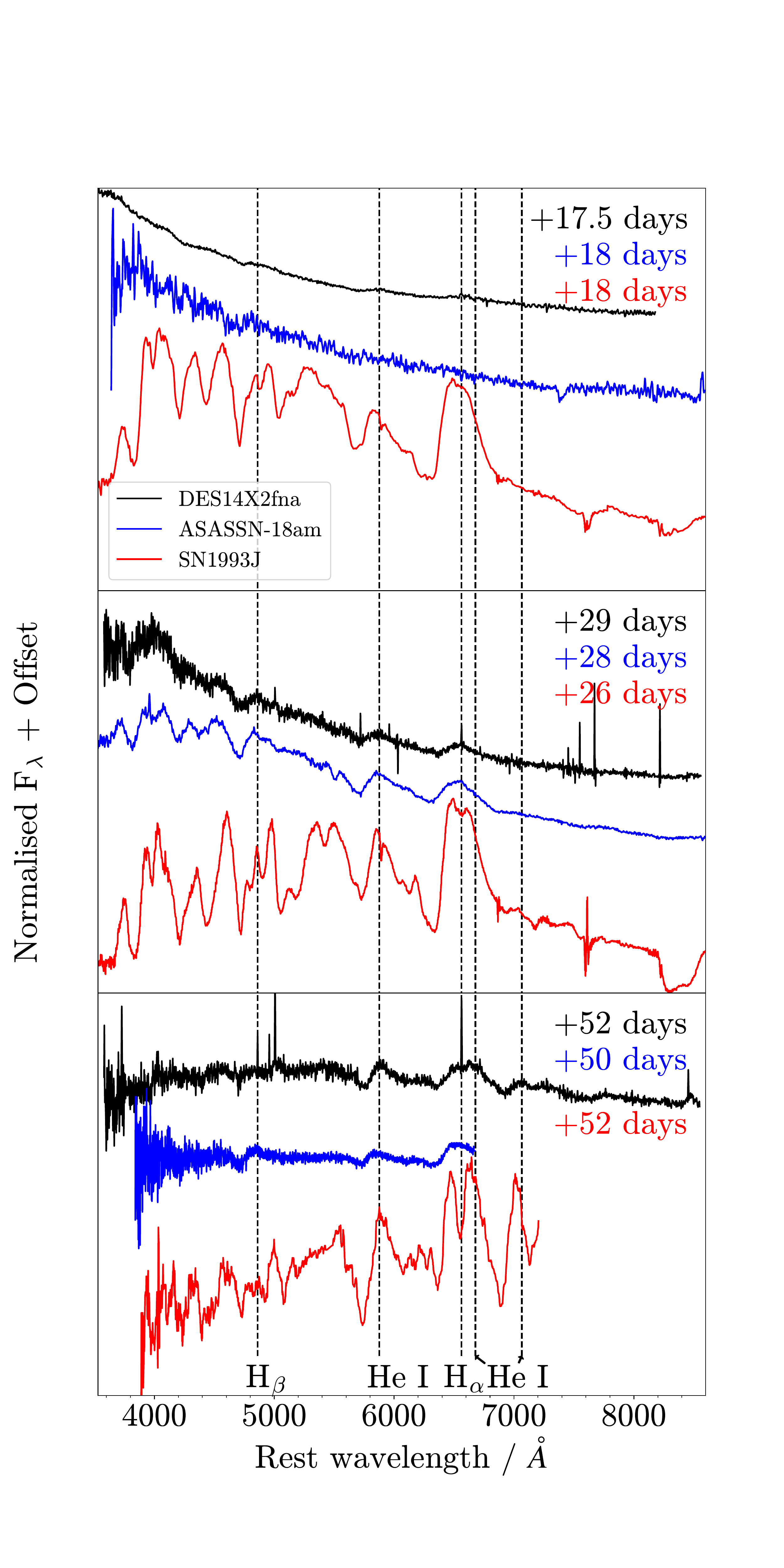}
\caption{Spectroscopy of DES14X2fna between 17.5 and 52 days after explosion, alongside spectra of ASASSN-18am and SN~1993J. Spectra have been arbitrarily shifted for presentation and corrected for Milky Way reddening using the extinction model of \citet{extinction}. Phases displayed in top right are with respect to explosion.} 
\label{spectra_comp}
\end{figure}

Fig.~\ref{spectra_comp} shows a comparison of the spectra of DES14X2fna with the similarly luminous SN IIb ASASSN-18am and the \lq prototypical\rq SN IIb SN~1993J. While DES14X2fna shows similar hydrogen and helium features to these objects, there are significant differences in the spectral evolution. Temperatures inferred from our black body fits in the lower panel of Fig.~\ref{BL} show that DES14X2fna and ASASSN-18am are significantly hotter than SN~1993J until $\sim$50 days after explosion, and this is reflected in the continuum.

After $\sim17.5$ days, DES14X2fna and ASASSN-18am have only subtle features and are dominated by a blue continuum while SN~1993J has clear features. SN~1993J displays H$\alpha$ and \ion{He}{I} 6678 lines which overlap to create a broader feature and an absorption feature appearing to correspond to H$\beta$, as well as \ion{He}{I} 5876 features.

After 29 days, DES14X2fna is still hotter than SN~1993J and comparable to ASASSN-18am. DES14X2fna again appears very similar to ASASSN-18am with both showing a broad H$\alpha$ feature as well as H$\beta$ and \ion{He}{I} 5876. For SN~1993J, at this epoch the broad feature at $\sim6600$ \AA\ shows dips around the wavelength of H$\alpha$, indicating that the contribution of H to this broad feature has reduced. After 52 days, the H$\beta$ feature in DES14X2fna is no longer visible though may still be contained in the noise. However, H$\alpha$ is still visible and \ion{He}{I} 6678 and 7065 lines are now apparent along with the \ion{He}{I} 5876 line seen previously. ASASSN-18am lacks wavelength coverage above $\sim$6700 \AA\ but still appears similar to DES14X2fna at this epoch, with H$\alpha$ and \ion{He}{I} 5876 features as well as a noticeable H$\beta$ feature. At this epoch, SN~1993J continues the trend of a dip in the broad feature at $\sim$6600 \AA\ around H$\alpha$, and now displays clear \ion{He}{I} 7065 lines.

Overall, the spectral evolution of these objects shows that DES14X2fna maintains its H envelope for far longer than SN~1993J, with H features visible in the spectra of DES14X2fna for far longer. This suggests that although the H envelope of the progenitor of DES14X2fna partially stripped, it still has a more massive H envelope than a typical SN IIb. This is also seen in ASASSN-18am, and \citet{asassn} discusses that this SN is also richer in H than other SNe IIb. Overall, DES14X2fna shows strong resemblance to ASASSN-18am over the epochs where we have spectroscopic coverage.

\section{Semi-analytic light curve modelling}
\label{mosfit}

Our analysis has shown significant differences between DES14X2fna and previously observed SNe IIb, with regard to both its luminosity and fast decline rate. We next consider the possible sources of luminosity of DES14X2fna by making use of the fitting code \textsc{MOSFiT} \citep{mosfit}. This uses a semi-analytic approach to light curve fitting, combining models for different sources to drive luminosity in the system with others to model diffusion through the SN ejecta to produce an observed luminosity over time. From these, model photometric light curves can be generated by assuming a modified black body SED. By using a Markov chain Monte Carlo (MCMC) approach to sample the parameter space for all of the models included, \textsc{MOSFiT} calculates the best fit light curve and parameters for a given model. We consider three different mechanisms included in \textsc{MOSFiT} to model the light curve of DES14X2fna:

\begin{enumerate}
    \item Nickel-cobalt decay: The canonical model of a SN IIb, using the treatment of the radioactive decay of $^{56}$Ni and $^{56}$Co from \citet{nicodecay}.
    \item Nickel-cobalt decay + CSM interaction: The nickel-cobalt decay model above along with extra luminosity resulting from interaction of the ejecta with a surrounding CSM, using the model from \citet{csm}. Although we do not observe the narrow spectral features associated with CSM interaction in DES14X2fna, this does not rule out the possibility that CSM interaction occured. Given a sufficiently low density CSM \citep{Reynolds20}, or if the SN ejecta collides with the CSM outside the broad-line forming region \citep{Arcavi17Nature}, CSM interaction can occur without clear narrow-line spectral features.
    \item Nickel-cobalt decay + magnetar: SN driven by a combination of nickel-cobalt decay, using the model outlined above, and a magnetar, using the model from \citet{nichollmag}. This is the proposed model for SN~1998bw in \citet{98bwmag}, which shows a close resemblance to DES14X2fna at peak. It is also suggested as an explanation for the light curve of ASASSN-18am in \citet{asassn}.
\end{enumerate}

\begin{figure*}
\begin{minipage}{\linewidth}
\centering
\includegraphics[width = \textwidth]{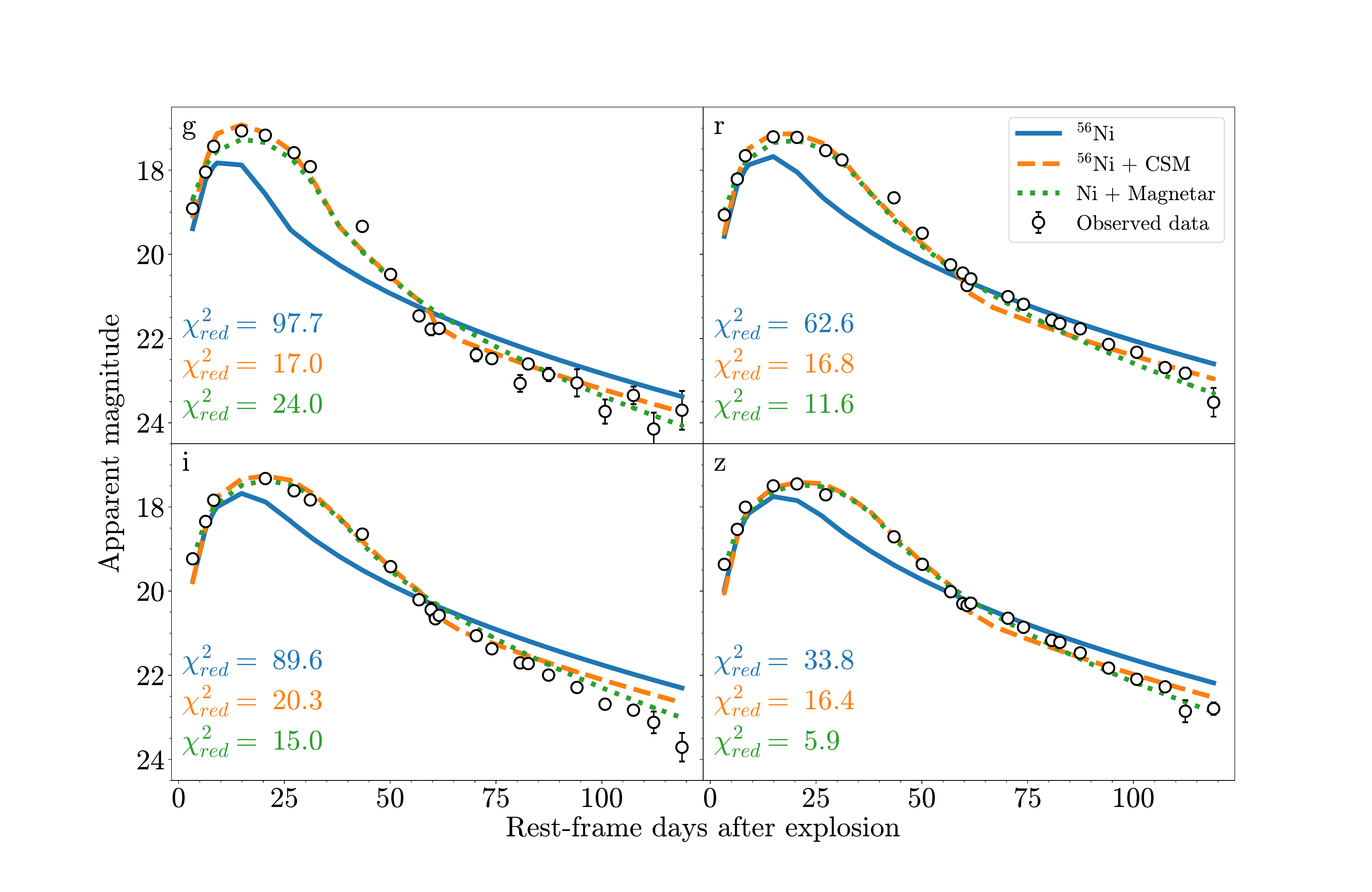}
\caption{The best-fit light curves to our observations for a variety of physical models after 10,000 MCMC iterations, along with the observed values in $griz$ bands.}
\label{model_compare}
\end{minipage}
\end{figure*}

For each of these models, we run an MCMC-based MOSFiT simulation comprising 1000 walkers for 10,000 iterations. For each walker, MOSFiT calculates model photometry at each epoch and compares with observed photometry. For each walker, we use model and observed photometry to calculate the reduced $\chi^2$ value, $\chi^2_\text{red}$ and use this to identify the best-fit model to observations. For these model fits, we have assumed a fixed opacity to UVOIR radiation $\kappa$=0.07 cm$^2$ g$^{-1}$, following \citet{Prentice2019}, as well as a fixed opacity to gamma rays produced in $^{56}$Ni decay of $\kappa_{\gamma,\text{Ni}}$=0.03 cm$^2$ g$^{-1}$ \citep{98bwmag}. Corner plots of each model fit are shown in Appendix \ref{appendix:a} (online supplementary material). Fig. \ref{model_compare} shows the best-fit light curve for each of these methods in DES $griz$ bands, along with the reduced $\chi^2$ value for each of these fits in each band. 

It is clear that nickel-cobalt decay alone is unable to accurately model the light curve of DES14X2fna, with an overall $\chi^2_\text{red} =$~58.2. This model predicts a narrower peak in $r$-band and a slower tail decline rate, especially in higher-wavelength bands than our observations. Additionally, this model overall underestimates luminosity in $r$-band and overestimates at late times in longer wavelength bands suggesting that it also fails to fit the high temperature and slow temperature decline of DES14X2fna. The full corner plot of this fit is presented in Appendix \ref{appendix:ni_plot} (online supplementary material), Fig \ref{ni_corner}, but in brief this model fails to give a reasonable value of ejecta mass and is consistent with a value below our chosen lower limit of 1 M$_{\odot}$. Prolonged H features in spectroscopy of DES14X2fna indicate a more massive hydrogen envelope than the prototypical SN IIb SN~1993J, suggesting that the ejecta mass cannot be substantially smaller than other objects - this motivated our choice of lower limit and and rules out such a low value.  This was obtained even when using a Gaussian prior with a mean and standard deviation of 3 M$_{\odot}$ and 0.3 M$_{\odot}$ respectively, selected arbitrarily to try and force the model to identify a minimum with a physically reasonable ejecta mass. Overall, this model is unable to match our observations and does not have realistic best-fit values for all parameters.

The addition of CSM-interaction to the nickel-cobalt decay model, however, significantly improves the quality of the model, with a reduced $\chi^2$ of 13.1. This model fits both the peak and the decline much more accurately, although the decline rate is still slower than observed in higher wavelength bands.  The best-fit ejecta mass is 1.51$^{+0.20}_{-0.18}$ M$_\odot$, relatively low for a SN IIb but still consistent with the sample in \citet{Prentice2019}. This model fits kinetic energy rather than ejecta velocity directly, but using the best-fit kinetic energy of 1.05$^{+0.10}_{-0.09}\times10^{51}$ erg s$^{-1}$ along with the ejecta mass, we estimate an ejecta velocity of $\sim$10,800 km s$^{-1}$, comparable to our velocities inferred from spectroscopy. This model is able to match most features of our observed photometry and the physical parameter values are consistent with properties expected for a SN IIb.

The addition of a magnetar to nickel-cobalt decay further improves the quality of the fit, with a reduced $\chi^2$ of 10.5. Compared with the CSM model, this model matches the observed decline rate of DES14X2fna in the tail more accurately. The best-fit ejecta mass is 2.78$\pm$0.50 M$_\odot$, far more typical for a SN IIb compared with the \citet{Prentice2019} sample. For this fit, we have assumed a fixed value of $\kappa_{\gamma,\text{mag}}$ = 0.01 cm$^2$ g$^{-1}$. We did experiment with leaving this as a free parameter but the MCMC favoured a value below the range of expected values for a magnetar, between 10$^{-2}$ and 10$^{6}$ cm$^2$ g$^{-1}$ \citep{kotera13}. As a result, we fix this to the lowest value considered reasonable.

\begin{figure}
\begin{minipage}{\linewidth}
\centering
\includegraphics[width = \textwidth]{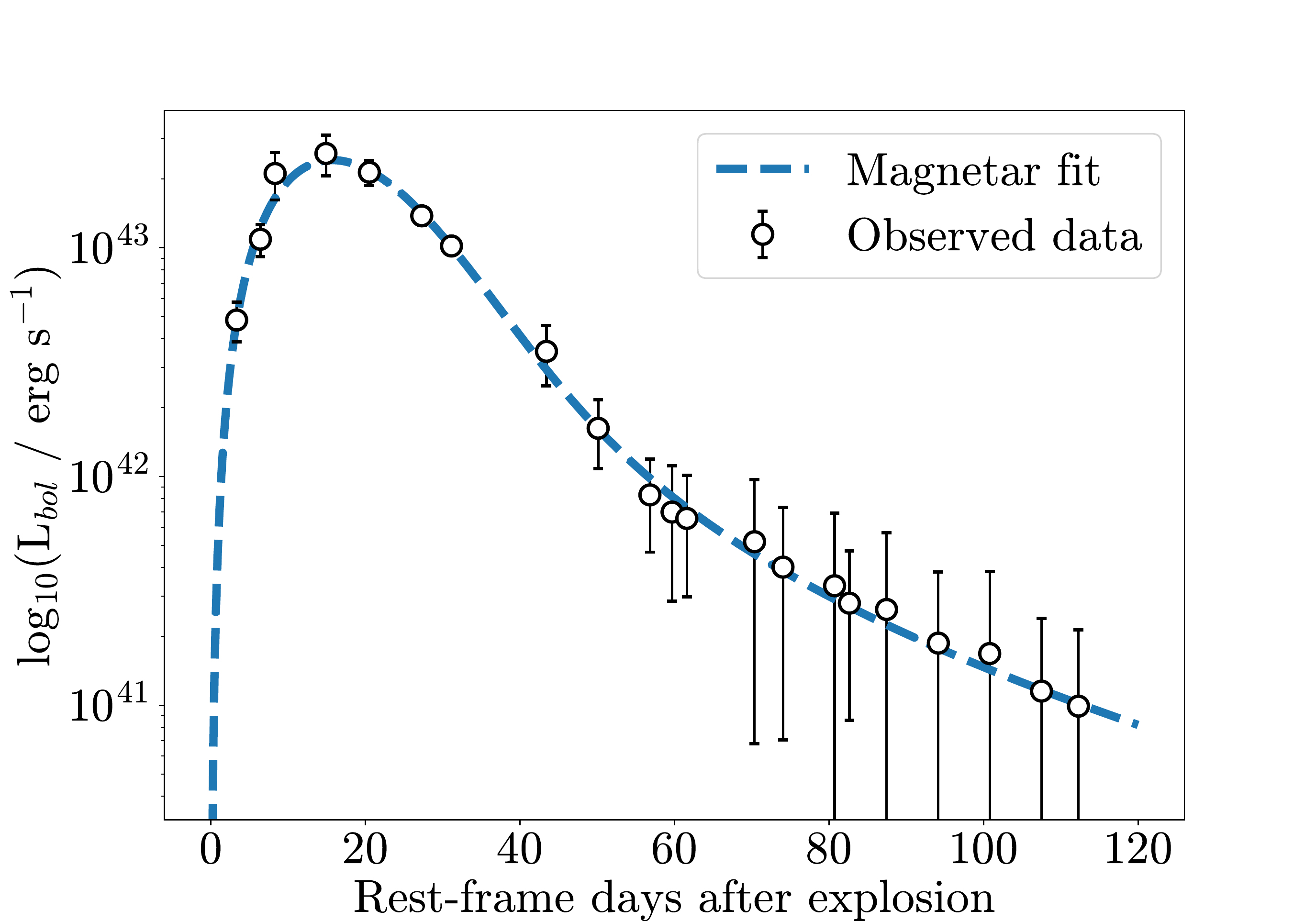}
\caption{Bolometric luminosities of DES14X2fna calculated from black body fits to observed photometry along with a magnetar fit to this data based on the model detailed in \citet{cosimo13}.}
\label{bol_magnetar_fit}
\end{minipage}
\end{figure}

As an independent check, we also considered an alternative semi-analytic magnetar code based on the model outlined in \citet{cosimo13}, hereafter \citetalias{cosimo13}. Unlike MOSFiT, \citetalias{cosimo13} fits directly to bolometric luminosities rather than observed photometry - as such, we utilise our black-body estimated luminosities from Section \ref{bol_lum}. For this fit, we have again assumed a fixed value of $\kappa_{\gamma,\text{mag}}$ = 0.01 cm$^2$ g$^{-1}$. Our fit to this data using \citetalias{cosimo13} is shown in Fig \ref{bol_magnetar_fit} - overall, this model does provide us with a good fit to our data both around peak and in the tail.

As previously noted, the $r$-band light curve of DES14X2fna shows strong resemblance peak to that of SN~1998bw (SN~Ic-BL), which \citet{98bwmag} finds can be accurately fit with a combination of a magnetar and nickel-cobalt decay. Our MOSFiT and \citetalias{cosimo13} magnetar fits indicate that a similar mechanism could power DES14X2fna. Table~\ref{magnetar_props} shows the best-fit properties of our magnetar model fit for DES14X2fna along with those of SN~1998bw. For both models, the predicted initial ejecta velocity v$_\text{ejecta}$ is consistent with our velocity measurements from spectroscopy and the ejecta mass is consistent with the SNe IIb population in \citet{Prentice2019}. Comparing DES14X2fna to SN~1998bw, there are differences of a factor of $\sim$2-2.5 for the estimated values of magnetar magnetic field strength and spin period. Both ejecta mass and velocity are lower for DES14X2fna although only by less than 1.5$\sigma$. One key difference between the two objects is that for DES14X2fna, the $^{56}$Ni mass is considerably less than SN~1998bw. The quoted $^{56}$Ni mass for \citetalias{cosimo13}, <0.0247 M$_\odot$, is a 1$\sigma$ upper limit from the fitting uncertainty - the parameter value itself is $\sim10^{-25}$. Under the assumption that DES14X2fna is powered in part by a magnetar, this demonstrates that the contribution of $^{56}$Ni decay to the light curve is minimal. The lower magnetic field strength and spin period for DES14X2fna compared to SN~1998bw have the effect of increasing the initial energy output of the magnetar and increasing the time scale over which it declines. In combination with the lower ejecta mass, this leads to DES14X2fna and SN~1998bw showing similar light curve peaks before DES14X2fna declines far more rapidly at later times.

\begin{table*}
\caption[]{The best fit properties of the combined magnetar and $^{56}$Ni fit to photometry of DES14X2fna using both MOSFiT and \citetalias{cosimo13}, along with the best-fit magnetar properties of SN~1998bw from \citet{98bwmag}.}
\begin{center}
\begin{tabular}{ |c|c|c|c|c|c|c| }
	\hline
	SN & B & P$_{\text{spin}}$& M$_{\text{ejecta}}$ & v$_\text{ejecta}$ & M$_{\text{Ni}}$ & $\kappa_{\gamma,\text{mag}}$  \\
	& ($10^{14}$ G) & (ms) & (M$_\odot$) & 10$^3$ km/s & (M$_\odot$) & (cm$^2$ g$^{-1}$) \\
	\hline
	DES14X2fna (MOSFiT) & 8.75$^{+5.81}_{-2.25}$ & 8.46$^{+1.94}_{-2.06}$ & 2.78$\pm$0.50 & 8.97$^{+0.66}_{-0.64}$ & (4.53$^{+6.78}_{-3.39}$)$\times10^{-3}$ & 0.01** \\
	DES14X2fna (\citetalias{cosimo13}) & 7.83$\pm$2.22 & 11.96$\pm$0.26 & 2.08$\pm$0.42 & 9.05$\pm$1.97 & <0.0247* & 0.01** \\
	SN~1998bw & 16.6$^{+2.1}_{-1.4}$ & 20.8$\pm0.8$ & 2.6$^{+0.5}_{-0.4}$ & 11.05$^{+1.50}_{-1.59}$ & 0.10$^{+0.03}_{-0.02}$ & 0.29$^{+0.19}_{-0.17}$ \\
    \hline
\end{tabular}
\end{center}
\begin{list}{}{}
\item \hspace{85pt} *: 1$\sigma$ upper limit
\item \hspace{85pt} **: Fixed value
\end{list}
\label{magnetar_props}
\end{table*}

\section{Discussion}
\label{discuss}

\subsection{Peak Luminosity}
\label{peak_nickel}

By studying both the $r$-band and bolometric light curves of DES14X2fna, we have demonstrated that it is considerably more luminous than any other SN IIb in our comparison sample. Under the assumption that this object follows the canonical $^{56}$Ni decay model for a SN IIb, the peak must be powered by a mass of $^{56}$Ni more than four times in excess of any other SN IIb. Using our bolometric light curve for DES14X2fna we can make some initial estimates of the mass of $^{56}$Ni synthesised in the explosion. Equation 3 of \citet{prentice_bol} from the commonly-used, analytically-derived `Arnett model' \citep{arnett82} allows us to estimate the $^{56}$Ni mass using the peak bolometric luminosity, L$_p$, and the rise time to peak bolometric luminosity, t$_p$. Using our values for these from section \ref{bol_lum} gives an estimate of M$_{Ni}$ = 1.02$\pm$0.20 M$_\odot$. Please note this is not a Gaussian uncertainty, rather a maximum possible deviation based on our uncertainty in explosion epoch, constrained only between the last non-detection and first detection. Equation 19 of \citet{k&k} proposes an updated relation between peak time and luminosity, with the effect of opacity and mixing encapsulated in a parameter $\beta$. Assuming that the proposed $\beta$ value for SNe IIb in \citet{k&k}, 0.82, is valid for DES14X2fna, we obtain M$_{Ni}$ = 0.633$^{+0.083}_{-0.079}$\ M$_\odot$. Again, this is not a Gaussian uncertainty, instead reflecting the maximum possible deviation due to uncertainty in explosion epoch. \citet{mass_distribution} finds SNe IIb and Ib $^{56}$Ni masses which typically vary from $\sim0.03 - 0.2$ and $\sim0.02 - 0.13$ M$_\odot$ using these methods respectively. DES14X2fna would require a $^{56}$Ni mass that is $\sim$ 4-5 times larger than is typical for SNe~IIb.

The breadth of the peak of the light curve of DES14X2fna also indicates an ejecta mass at the very least typical for a SN IIb. We can estimate the ejecta mass using equation 1 in \citet{Prentice2019}, based on the Arnett model,

\begin{equation}
    M_{ej} = \frac{1}{2}\Bigg(\frac{\beta c}{\kappa}\Bigg)\tau_m^2 v_{sc}
\end{equation}
where $\beta\approx13.7$ is a constant of integration, $\kappa$ is the opacity of the ejecta and $v_{sc}$ is a characteristic velocity of the ejecta. We again follow \citet{Prentice2019} in assuming $\kappa$ = 0.07 cm$^2$ g$^{-1}$ and $\tau_m$ = t$_p$, and use the velocity of H$\alpha$ around peak of $\simeq$9000 km s$^{-1}$ as an estimate of the characteristic ejecta velocity. This gives an estimate for ejecta mass of $M_{ej} = 2.35^{+1.11}_{-0.90}$ M$_\odot$, consistent with the mean and median $M_{ej}$ for SNe IIb in \citet{Prentice2019}, 2.7 ± 1.0 M$_\odot$ and $2.5^{+1.3}_{-0.7}$ M$_\odot$, respectively. This suggests that the light curve of DES14X2fna is consistent with a fairly typical SN IIb ejecta mass.

\subsection{Tail Decline}
\label{tail_decline}

As well as the high peak luminosity and broad peak of DES14X2fna, we have also demonstrated that DES14X2fna declines far more rapidly at late times than any other SN in our comparison sample. However, under the assumption that this object is powered by $^{56}$Ni decay, this creates a discrepancy with the very luminous peak of DES14X2fna.

In the canonical picture of a SN IIb, after any hydrogen present has recombined the light curve is driven by radioactive decay of $^{56}$Co, as any $^{56}$Ni produced will have already decayed. The source of optical luminosity at this stage is the deposition of $\gamma$-rays and positrons produced in the radioactive decay process. Under the assumption that $\gamma$-rays produced are fully trapped and all deposit their energy, the luminosity of the SN will decline at a rate of 0.98 mag (100d)$^{-1}$ \citep{trapped_decay_rate}. A decline rate faster than this indicates that $\gamma$-rays produced are not fully trapped and therefore some of their energy is not observed in the optical light curve. In $griz$ photometric bands, DES14X2fna has decline rates of 3.69 $\pm$ 0.48 mag (100d)$^{-1}$, 4.38 $\pm$ 0.10 mag (100d$^{-1}$), 5.21 $\pm$ 0.14 mag (100d)$^{-1}$ and 4.69 $\pm$ 0.10 mag (100d)$^{-1}$, far in excess of the rate expected for complete trapping. Such a fast decline rate indicates a very low ejecta mass, allowing most of the $\gamma$-rays produced to escape. However, as discussed in section \ref{peak_nickel}, if DES14X2fna were to be powered by $^{56}$Ni it would require a mass of $^{56}$Ni at least four times that of any other SN IIb observed and an ejecta mass at least typical for a SN IIb. This high mass of material would lead to significant trapping of the $\gamma$-rays produced in $^{56}$Co decay and therefore preclude the fast decline which is observed in DES14X2fna. Assuming $^{56}$Ni alone, DES14X2fna cannot have an ejecta mass typical for a SN IIb while also displaying a significantly faster decline rate.

We can also estimate $^{56}$Ni mass from the bolometric tail decline, which helps to reiterate the discrepancy between peak and decline rate for DES14X2fna. For a canonical SN IIb, in this phase $^{56}$Ni mass will have almost entirely decayed and the light curve is powered solely by $^{56}$Co. Assuming complete $\gamma$-ray trapping, in this phase the light curve will follow \citep{Jerkstrand2012},

\begin{equation}
    L(t) = 1.42\times10^{43}\frac{M_{Ni}}{M_\odot}(e^{-t/t_{Co}}-e^{-t/t_{Ni}}) \text{ erg s}^{-1},
\label{L0}
\end{equation}
where $L(t)$ is the luminosity, $M_{Ni}$ is the mass of $^{56}$Ni synthesised in the explosion and $t_{Co}$ and $t_{Ni}$ are the characteristic decay timescales for $^{56}$Co and $^{56}$Ni respectively. Assuming that other SNe follow this relation and have similar $\gamma$-ray deposition, SNe $^{56}$Ni masses will share the same ratio as bolometric luminosity at the same epoch. SN~1987A is commonly used for this analysis, although as it does not reach the linear decline phase of its light curve until after our last data for DES14X2fna, this method cannot be directly used. However, we can infer a bolometric luminosity at later times for DES14X2fna by extrapolating the linear decline in $\log(L_{bol})$. After 159.1 days since explosion SN~1987A has a UV-optical-infrared (UVOIR) bolometric luminosity of $\log_{10} (L_{\text{UVOIR}}$/ erg s$^{-1})$ =  41.381 \citep{87A}, whereas extrapolating the light curve of DES14X2fna gives a luminosity at this epoch of $\log_{10} (L_{bol}$/ erg s$^{-1})$ = 40.22$\pm$0.11. Note that the uncertainty for the SN~1987A luminosity is less than 0.02 dex, although no exact value is given. As a result, we consider its uncertainty to be negligible compared to DES14X2fna. The ratio of luminosities is (6.93$\pm$1.74) $\times10^{-2}$, leading to an estimate for $^{56}$Ni mass for DES14X2fna of (5.20$\pm$1.35) $\times10^{-3}$ M$_\odot$. The underlying assumption that these objects share similar $\gamma$-ray deposition is clearly not valid as DES14X2fna exhibits a very rapid decline not seen in SN~1987A, so this value is only a lower limit on the $^{56}$Ni mass produced. While only a lower limit, the fact that this value is nearly two orders of magnitude less than the mass of $^{56}$Ni predicted from peak luminosity demonstrates the inconsistency between peak and tail of DES14X2fna.

It is also possible to take into account the effect of incomplete trapping of $\gamma$-rays when estimating the $^{56}$Ni mass from the bolometric decline. \citet{clocchiatti} proposes a simple model to take this into account, also given in equation 2 of \citet{terreran_eq}. This is the most logical method to estimate $^{56}$Ni mass for DES14X2fna given its fast decline rate. However, we find that this model is unable to constrain the $^{56}$Ni mass of DES14X2fna despite working for every other object in our SNe IIb comparison sample. The reason for this appears to be that DES14X2fna declines too fast to be well fit by this model. Equation 7 of \citet{clocchiatti},

\begin{equation}
    \frac{d}{dt}m(t) = \frac{d}{dt}m_0(t) + \frac{5\log_{10}(e)}{e^{(T_0/t)^2}-1}\frac{T_0^2}{t^3}
\label{m_decline}
\end{equation}
gives an expression for the decline rate in magnitude for $^{56}$Co decay, where $\frac{d}{dt}m_0(t)$ is the decline rate for fully trapped $^{56}$Co decay, 0.98 mag (100d)$^{-1}$, $T_0$ is the characteristic decay time for the $\gamma$-ray optical depth and the complete second term corresponds to the shift in decline rate as a result of incomplete trapping, $\frac{d}{dt}m_1(t)$. If we are to assume a fixed value of t, the second term increases with decreasing $T_0$ and is maximised in the limit $T_0$ tends to zero. Applying this, the maximum value of the second term is,

\begin{equation*}
    \lim_{x \to 0} \frac{d}{dt}m_{1}(t) = \frac{5\log_{10}(e)}{t}.
\label{m_decline}
\end{equation*}

At the start of the tail after 60 days, this term is 3.62 mag (100d)$^{-1}$ giving a maximum decline rate of 4.60 mag (100d)$^{-1}$, which does not rule out the fast decline rate of DES14X2fna at this epoch. However, at later times the maximum decline rate decreases - for example, 100 days after explosion the maximum possible decline rate is 3.15 mag (100d)$^{-1}$ which is more than 1 mag (100d)$^{-1}$ less than the observed decline rate of DES14X2fna at this epoch. This discrepancy only increases at later stages of our observed light curve. Overall, this suggests that DES14X2fna declines too rapidly to be explained by $^{56}$Ni decay alone.

\subsection{Power source}

Throughout this paper, we have demonstrated that DES14X2fna has very different properties when compared with other SNe IIb. DES14X2fna reaches a peak absolute magnitude in $r$-band of -19.36, comparable only to ASASSN-18am and nearly 1 mag brighter than any other SN IIb. DES14X2fna also displays a relatively broad light curve peak. As discussed in section \ref{peak_nickel}, this is indicative of a very high $^{56}$Ni mass and an at least comparable ejecta mass when compared to other SNe IIb. However, as discussed in section \ref{tail_decline}, these $^{56}$Ni and ejecta masses are not consistent with the very fast post-peak decline observed across all photometric bands of DES14X2fna. DES14X2fna appears to decline too rapidly for this to be explained simply by incomplete trapping of $\gamma$-rays produced in $^{56}$Co decay. In section \ref{mosfit}, we consider a semi-analytic model of $^{56}$Ni decay. This model is unable to accurately model the light curve of DES14X2fna, predicting a narrower peak in $r$-band and a slower tail decline rate, especially in higher-wavelength bands, than our observations. This model is also unable to reproduce the temperature of DES14X2fna, underestimating luminosity at peak in shorter wavelength bands and overestimating at longer wavelengths at later times. Based on the contradictions between the peak and decline of the light curve of DES14X2fna as well as the inability of the semi-analytic model to fit observed photometry, we rule out the possibility that DES14X2fna can be powered by $^{56}$Ni decay alone.

Having established that DES14X2fna requires an additional source of luminosity besides $^{56}$Ni, the question to address is what that source of luminosity might be. One possibility is that interaction with a surrounding CSM helps to power the light curve. Such a model allows for a faster decline than $^{56}$Ni decay alone as the decline rate is no longer limited by the decay timescale of $^{56}$Co - a decrease in the luminosity from CSM interaction can lead to a much more rapid decline \citep{Taubenberger19}. There is precedent for SNe IIb which also show signs of interaction. SN~2018gjx \citep{2018gjx} was an object which spectroscopically resembled a SN IIb before going on to show narrow spectral features typical of a SN Ibn after $\sim$40 days, although DES14X2fna is two mags brighter at peak and does not show narrow spectral features during our spectroscopic coverage so is not directly comparable. Our semi-analytic fits using MOSFiT suggest that CSM interaction combined with $^{56}$Ni decay could power the light curve of DES14X2fna. Although we do not see the narrow spectral features associated with CSM interaction in spectroscopy of DES14X2fna, as previously mentioned this does not rule out that it occurred - interaction has been invoked for other objects which no not display narrow lines (e.g. SN~2016gsd, \citealt{Reynolds20}; iPTF14hls \citealt{Arcavi17Nature}).

We have also shown that the $r$-band peak of DES14X2fna shows close resemblance to the $R$-band peak of SN~1998bw, potentially suggesting that they share a source of luminosity at this phase of the light curve. However, DES14X2fna is also seen to decline far more rapidly than SN~1998bw at late times. If they are to share a source of luminosity, this source must be capable of powering the peak before quickly decreasing in luminosity. \citet{98bwmag} finds that the bolometric light curve of SN~1998bw is likely powered by a combination of $^{56}$Ni decay and a magnetar, with this model favoured over the traditional two-component model for SNe Ic-BL as it can explain the late time shift in decline rate. Crucially, \citet{98bwmag} finds that the magnetar dominates the light curve only during peak and at very late times, with the post-peak decline dominated by $^{56}$Ni decay. The very luminous peak and fast decline of DES14X2fna could therefore potentially be explained by a magnetar. Our semi-analytic model fits find that a combination of a magnetar and $^{56}$Ni decay is able to fit the light curve of DES14X2fna far better than $^{56}$Ni decay, accurately modelling the peak and decline rate of DES14X2fna.

The similarity of DES14X2fna to SN~1998bw and the good fit to our observed photometry given by the combined magnetar and $^{56}$Ni model provides evidence that DES14X2fna could also partially powered by a magnetar. This model also demonstrates that $^{56}$Ni makes a minimal contribution to the light curve. This helps to explain the differences between DES14X2fna and SN~1998bw. As shown in \citet{98bwmag}, the light curve of SN~1998bw is dominated by a magnetar only around peak and at very late times, far later than photometry is available for DES14X2fna. At intermediate times, the light curve is dominated by $^{56}$Co decay. In the case of DES14X2fna, it may be that the light curve is still dominated by a magnetar at intermediate times, which would allow for a faster decline rate than is typically associated with $^{56}$Co decay. This would explain why DES14X2fna declines far more rapidly after peak than SN~1998bw. Overall, a magnetar model provides a good fit to our observations of DES14X2fna.

An additional possibility to consider is whether the rapid decline of DES14X2fna could result from the onset of dust formation. This process has been invoked to explain fast declines observed in SNe Ibn such as SN~2006jc \citep{SN2006jc}. For this object, the optical light curve was seen to flatten $\sim$30-50 days after maximum before a sudden increase in optical decline rate was accompanied by an increase in NIR luminosity as a result of dust reprocessing. In the absence of NIR observations for DES14X2fna we cannot be sure whether this might have occurred, although it is important to note that SN~2006jc was an object which showed strong interaction and had a much faster evolution than DES14X2fna - even during the flattening phase, it still declined faster than DES14X2fna at optical wavelengths. As a result, we consider the possibility that this decline rate is due to dust formation unlikely although we cannot rule it out.

\subsection{DES14X2fna as a photometric contaminant}
\label{contaminant}

DES14X2fna has a very high luminosity at peak, reaching a peak M$_{r}$ $\simeq$ -19.3, and therefore into the range of peak luminosities expected for type Ia SNe. While the H and He features of DES14X2fna clearly distinguish it from a SN Ia, it is a possibility that a similar object could be mistakenly included in a photometric sample of SNe Ia in the absence of spectroscopy. To assess this, we use the SNe Ia light curve template fitter SALT2 \citep{SALT2} to fit our observed photometry using the Python package \textsc{sncosmo}. SALT2 calculates the stretch and colour parameters for a SN Ia light curve, $x_1$ and $c$. For the DES cosmological sample of SNe Ia, only SNe with -0.3<$c$<0.3 and -3<$x_1$<3 are included. Fitting the observed light curve between -15 and +45 days relative to explosion gives a poor quality fit, particularly in higher wavelength bands, with $\chi^2_{red} = 6041$. This fit also has a value of $x_1=-3.08$ and $c=-1.20$, meaning that it would not pass the cuts to be included in a cosmological sample. However, the large $\chi^2_{red}$ value obtained also results partly from the fact that DES14X2fna has very high signal-to-noise data. If a similar object were observed at a higher redshift, it is possible that it could be misclassified as a SN Ia.

To assess the likelihood of an object similar to DES14X2fna to act as a photometric contaminant, we simulate the object based on SED templates from Hounsell et al. (in prep.) over the 5 years of DES using \textsc{SNANA} \citep{SNANA}. These templates can extend into the UV based on extrapolation of the optical SED but effects such as line blanketing mean that these are not reliable. As a result, we simulate objects up to z<0.59 meaning that the effective wavelength of DES $g$-band never falls below 3000 \AA. In total, 23,870 DES14X2fna-like events are simulated at different redshifts in DES up to the this limit. Each of these synthetic light curves is fit using \textsc{SALT2}, a standard template fitter for SNe Ia light curves \citep{SALT2}, and the recurrent neural network (RNN)-based photometric classifier \textsc{SuperNNova} \citep{supernnova} is used to estimate the probability that each light curve corresponds to a SN Ia, hereafter referred to as P$_\text{Ia}$. The model used for classification was trained on a set of templates which did not include DES14X2fna and so has not previously seen an object with these properties. To identify the synthetic light curves with the potential to be mistakenly included in cosmological samples of SNe Ia, we apply a number of cuts:
\begin{enumerate}
    \item Select only SNe with valid SALT2 fits: This identifies SNe with light curves with the potential to be included in the DES 3-year cosmological sample based on the cuts outlined in \citet{Brout_2019}. Any objects outside the redshift range of [0.05,1.2], with a Milky Way E(B-V) > 0.25, without data both before and after peak or without data points with a signal-to-noise ratio > 5 in at least two bands are excluded from the SALT2 fits entirely, leaving a total of 18,520 objects.
    \item Select only SNe with -0.3<$c$<0.3 and -3<$x_1$<3 based on SALT2 fits: As previously mentioned, this cut is used in \citet{Brout_2019} to select the cosmological sample. Performing this cut leaves a total of 15,101 objects in the sample.
    \item Select only SNe with a high value of P$_\text{Ia}$: This identifies only the synthetic SNe with a high probability of being classified as a SN Ia by \textsc{SuperNNova}. Note that \textsc{SuperNNova} has two models, one for photometric classification with a host spectroscopic redshift and one without. Here we consider only the model for classification with a redshift as objects without host redshifts would not be included in a cosmological sample. This confidence threshold for inclusion is arbitrary - a lower threshold will give a larger sample but increases the likelihood of source being included incorrectly. Using a high threshold of P$_\text{Ia}$ > 0.9 gives a total of 270 synthetic SNe which would be misclassified as SNe Ia, 1.1 per cent of the sample. Lowering this threshold to 0.8 increases the number of misclassifications to 372, 1.6 per cent of the sample. At a threshold of P$_\text{Ia}$ > 0.5, where an object is simply more likely than not to be a SN Ia, there are 565 misclassifications corresponding to 2.4 per cent of the sample.
    
\end{enumerate}

Overall, in our simulations a DES14X2fna-like object is misclassified as a SN Ia in $\sim$1.1-2.4 per cent cases depending on the threshold used for P$_\text{Ia}$. While rare, it does show that such an object does have the potential to contaminate a cosmological sample. Analysing the light curves of the synthetic SNe which were classified shows that this generally occurs when the fast linear decline of DES14X2fna is undetected, due to either dropping below the detection limit of the survey or the end of the observing season, or is observed with very low signal-to-noise. This suggests that DES14X2fna is difficult to differentiate from a SN Ia based on photometry around peak, but can be differentiated from its late time decline. This is to be expected given DES14X2fna does not show the second peak typical for SNe Ia and has a very fast tail decline. 

Given that DES14X2fna is near unique, it is clear that this type is rare with a rate only a small fraction of the rate of SNe Ia. In addition to this, only a small fraction of similar events will be misclassified as SNe Ia. While DES14X2fna is an interesting case of a potential contaminant, in practice the effect that these objects have on cosmological analyses is unlikely to be significant.

\section{Conclusions}
\label{conclusions}

DES14X2fna is an unusual SN IIb with both a very high peak luminosity and very fast late-time decline compared to other SNe IIb. At peak, this SN reaches an $r$-band absolute magnitude of -19.37 $\pm$ 0.05, comparable to the recently discovered very luminous SN IIb ASASN-18am and 0.88 mag brighter than any other SN IIb. The light curve is also fairly broad for a SN IIb. Our main conclusions are as follow:

\begin{enumerate}
    \item The light curve of DES14X2fna cannot be explained by $^{56}$Ni decay alone. The peak luminosity and peak time of DES14X2fna indicate a $^{56}$Ni mass synthesised in the explosion more than four times greater than observed in any other SNe IIb in \citet{k&k} and an ejecta mass of $2.20^{+1.08}_{-0.87}$ M$_\odot$ which is consistent with other SNe IIb in \citet{Prentice2019}. However, DES14X2fna declines at more than three times the rate expected for fully-trapped $^{56}$Co decay in $g$-band and more than four times in $riz$-bands. This decline is more than 1 mag 100 d$^{-1}$ faster than any other SN IIb in the tail.
    \item DES14X2fna displays signs of H for far longer than the prototypical SN IIb SN~1993J, indicating it has a more massive H envelope than a typical SN IIb. \citet{asassn} obtains a similar finding for ASASSN-18am.
    \item Our semi-analytic $^{56}$Ni decay models using MOSFiT are unable to fit the peak and the fast tail decline of the light curve of DES14X2fna.
    \item DES14X2fna declines too rapidly at late times to be fit by the treatment for incomplete trapping of $\gamma$-rays produced in $^{56}$Co decay in \citet{clocchiatti}, providing further evidence that an additional source of luminosity is required.
    \item The addition of CSM interaction to the $^{56}$Ni decay model provides a significantly better fit to the observed photometry of DES14X2fna than $^{56}$Ni decay alone, indicating that it could in part power this object.
    \item The light curve of DES14X2fna is well fit by a magnetar model. DES14X2fna also shows resemblance to SN~1998bw around peak, which is well fit by a magnetar model in \citet{98bwmag}.
    \item Based on simulations of DES14X2fna in SNANA using templates from Hounsell et al. (in prep.) in DES, we find that such an object is misclassified as a SN Ia suitable for cosmology in $\sim$1.1-2.4 per cent cases depending on the probability threshold used. This typically occurs when the fast decline of such an object is not observed with sufficient signal-to-noise. However, given the rarity of such events they are unlikely to act as a significant contaminant to cosmological samples.
\end{enumerate}

\appendix

\section{MOSFiT Details (online supplementary material)}
\label{appendix:a}

\subsection{Nickel Decay}

\subsubsection{Prior}
\label{prior1}

Table \ref{ni_prior} gives the details of the priors used on parameter values in our MOSFiT simulation for the $^{56}$Ni decay model. Flat and log-flat values were left at their default values, Gaussian priors were selected based on physical observations which can be used to infer ejecta mass and velocity. These priors were selected to restrict the parameter space to physically reasonable solutions. MOSFiT models the photosphere as expanding and cooling with the ejecta before receding at a constant final temperature as described in \citet{nichollmag} - this is the parameter T$_\text{min}$. We have marginalised over nuisance parameters including n$_\text{H,host}$, the line-of-sight H number density of the host used to calculate extinction which is not constrained by our data, and the white-noise variance (see \citet{nichollmag}), the additional magnitude uncertainty required to give a reduced $\chi^2$ equal to one.

\begin{table*}
\caption[]{Details of priors selected for $^{56}$Ni decay model for each parameter, detailing minimum and maximum values, prior type and mean and standard deviation in the case of a Gaussian prior.}
\begin{center}
\begin{tabular}{ |c|c|c|c|c|c|c|c| }
	\hline
	Parameter & Description & Unit & Minimum Value & Maximum Value & Prior & $\mu$ & $\sigma$ \\
	\hline
	f$_\text{Ni}$ & $^{56}$Ni fraction of ejecta &  -- & 10$^{-3}$ & 1 & flat & -- & -- \\
	T$_\text{min}$ & Minimum photosphere temperature & K & 10$^{3}$ & 10$^{5}$ & log-flat & -- & -- \\
	M$_\text{ejecta}$ & Ejecta mass & M$_\odot$ & 1 & 4 & Gaussian & 3 & 0.3 \\
	v$_\text{ejecta}$ & Ejecta velocity & km s$^{-1}$ & 6000 & 18000 & Gaussian & 9000 & 1000 \\
    \hline
\end{tabular}
\end{center}
\label{ni_prior}
\end{table*}

\subsubsection{Corner Plot}
\label{appendix:ni_plot}

\begin{figure*}
\begin{minipage}{\linewidth}
\centering
\includegraphics[width = \textwidth]{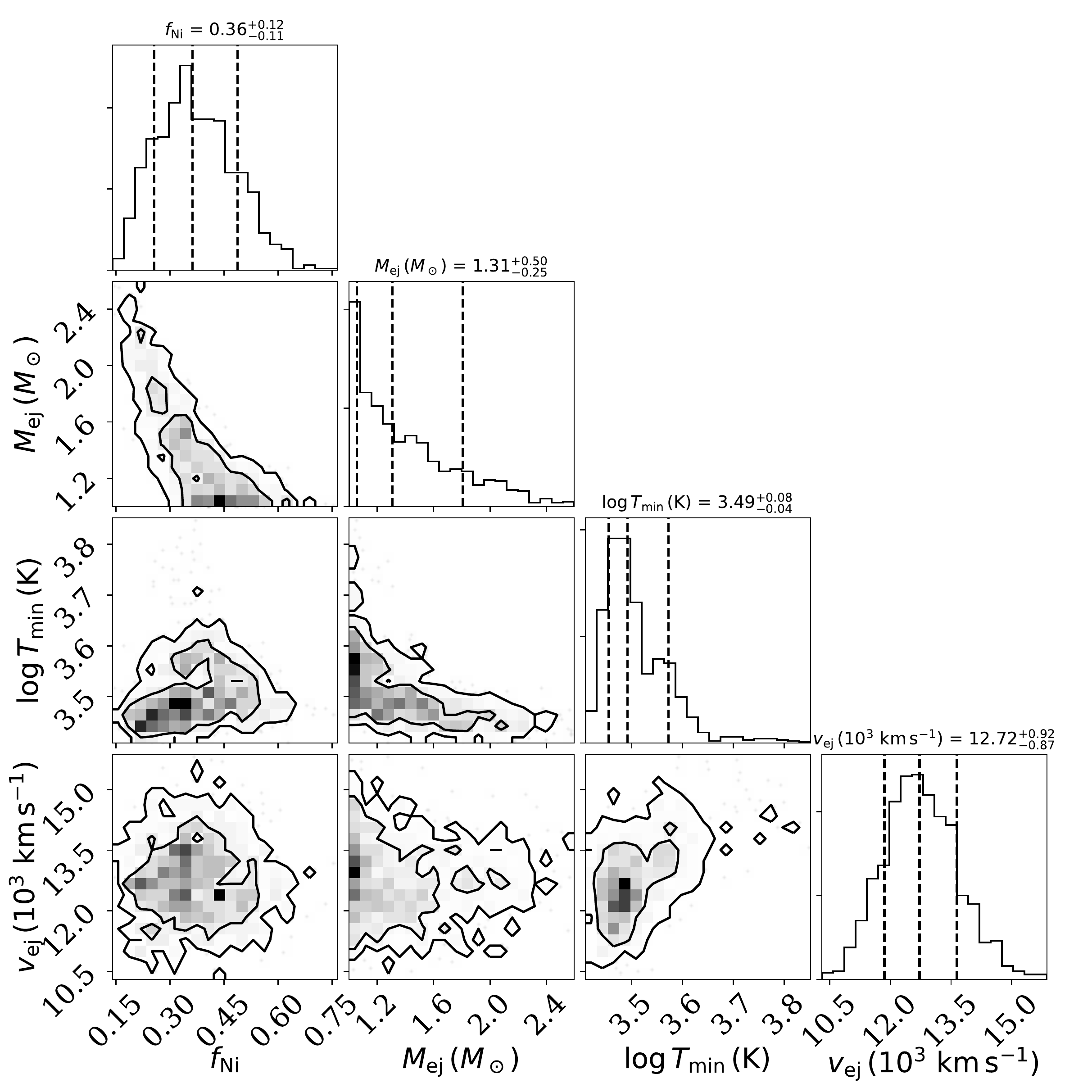}
\caption{Corner plot of the $^{56}$Ni decay model fit to photometry of DES14X2fna using MOSFiT, showing the best-fit parameter values and their correlations and degeneracies. Dashed lines represent 16th, 50th and 84th percentiles and contours represent 1$\sigma$ and 2$\sigma$ confidence levels.}
\label{ni_corner}
\end{minipage}
\end{figure*}

Figure \ref{ni_corner} shows the corner plot detailing best-fit parameter values and their correlations with each other for the $^{56}$Ni decay model.

\subsection{Nickel Decay and CSM Interaction}

\subsubsection{Prior}

Table \ref{csmni_prior} gives the details of the priors used on parameter values in our MOSFiT simulation for the combined $^{56}$Ni decay and CSM interaction model. As mentioned in section \ref{prior1}, flat and log-flat values were left at their default values and Gaussian priors were selected to restrict the parameter space to physically reasonable solutions based on our observations.

\begin{table*}
\caption[]{Details of priors selected for combined $^{56}$Ni decay and CSM interaction model for each parameter, detailing minimum and maximum values, prior type and mean and standard deviation in the case of a Gaussian prior.}
\begin{center}
\begin{tabular}{ |c|c|c|c|c|c|c|c| }
	\hline
	Parameter & Description & Unit & Minimum Value & Maximum Value & Prior & $\mu$ & $\sigma$ \\
	\hline
	f$_\text{Ni}$ & $^{56}$Ni fraction of ejecta & -- & 10$^{-3}$ & 1 & flat & -- & -- \\
	T$_\text{min}$ & Minimum photosphere temperature & K & 10$^{3}$ & 10$^{5}$ & log-flat & -- & -- \\
	M$_\text{CSM}$ & Mass of CSM & M$_\odot$ & 10$^{-2}$ & 10 & Gaussian & 0.3 & 0.1 \\
	M$_\text{ejecta}$ & Mass of ejecta & M$_\odot$ & 1 & 4 & Gaussian & 2 & 0.5 \\
	E$_K$ & Kinetic energy of ejecta & 10$^{51}$ erg & 0.5 & 1.5 & Gaussian & 1.03 & 0.1 \\
	$\rho$ & Density of CSM & g cm$^{-2}$ & 10$^{-15}$ & 10$^{-11}$ & log-flat & -- & -- \\
    \hline
\end{tabular}
\end{center}
\label{csmni_prior}
\end{table*}


\subsubsection{Corner Plot}

\begin{figure*}
\begin{minipage}{\linewidth}
\centering
\includegraphics[width = \textwidth]{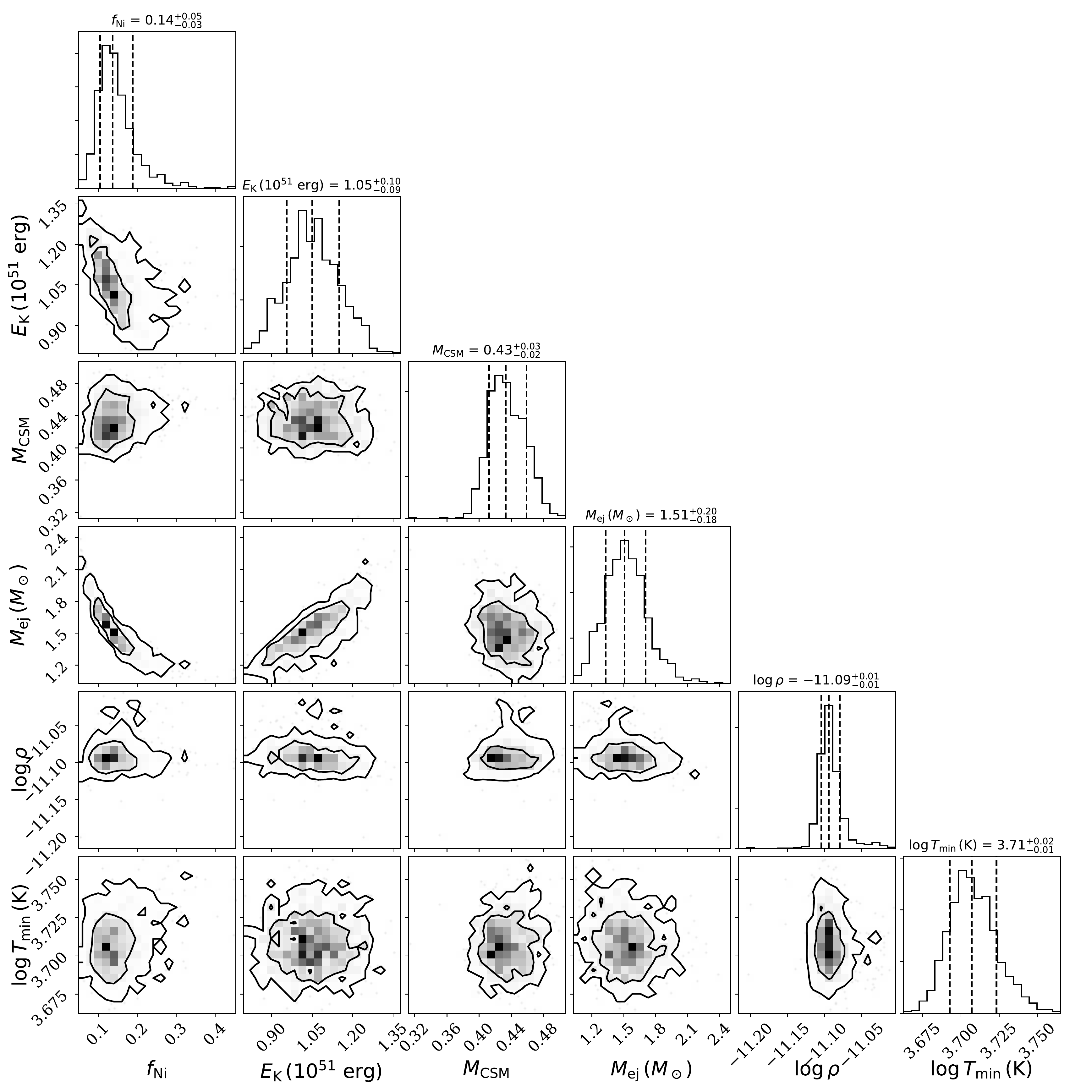}
\caption{Corner plot of the combined $^{56}$Ni decay and CSM interaction model fit to photometry of DES14X2fna using MOSFiT, showing the best-fit parameter values and their correlations and degeneracies. Dashed lines represent 16th, 50th and 84th percentiles and contours represent 1$\sigma$ and 2$\sigma$ confidence levels.}
\label{csmni_corner}
\end{minipage}
\end{figure*}

Figure \ref{csmni_corner} shows the corner plot detailing best-fit parameter values and their correlations with each other for the combined $^{56}$Ni decay and CSM interaction model.

\subsection{Magnetar}

\subsubsection{Prior}

Table \ref{csmni_prior} gives the details of the priors used on parameter values in our MOSFiT simulation for the magnetar model. As mentioned in section \ref{prior1}, most flat and log-flat priors were left at their default values and Gaussian priors were selected to restrict the parameter space to physically reasonable solutions based on our observations. We have raised the minimum value of P from the default of 1 ms to 5 ms as with the lower threshold some walkers were ending in a local minimum with an unphysically small ejecta velocity. We fix the neutron star mass at 1.4 M$_\odot$.

\begin{table*}
\caption[]{Details of priors selected for combined $^{56}$Ni decay and CSM interaction model for each parameter, detailing minimum and maximum values, prior type and mean and standard deviation in the case of a Gaussian prior.}
\begin{center}
\begin{tabular}{ |c|c|c|c|c|c|c|c| }
	\hline
	Parameter & Description & Unit & Minimum Value & Maximum Value & Prior & $\mu$ & $\sigma$ \\
	\hline
	f$_\text{Ni}$ & $^{56}$Ni fraction of ejecta & -- & 10$^{-3}$ & 1 & flat & -- & -- \\
	T$_\text{min}$ & Minimum photosphere temperature & K & 10$^{3}$ & 10$^{5}$ & log-flat & -- & -- \\
	B & Magnetar magnetic field strength & 10$^{14}$G & 0.1 & 20 & flat & -- & -- \\
	P$_\text{spin}$ & Spin-down time-scale of magnetar & ms & 5 & 20 & flat & -- & -- \\
	M$_\text{ejecta}$ & Ejecta mass & M$_\odot$ & 0.1 & 10 & Gaussian & 2 & 0.5 \\
	v$_\text{ejecta}$ & Ejecta velocity & km s$^{-1}$ & 1000 & 20000 & Gaussian & 9000 & 1000 \\
    \hline
\end{tabular}
\end{center}
\label{magni_prior}
\end{table*}

\subsubsection{Corner Plot}

\begin{figure*}
\begin{minipage}{\linewidth}
\centering
\includegraphics[width = \textwidth]{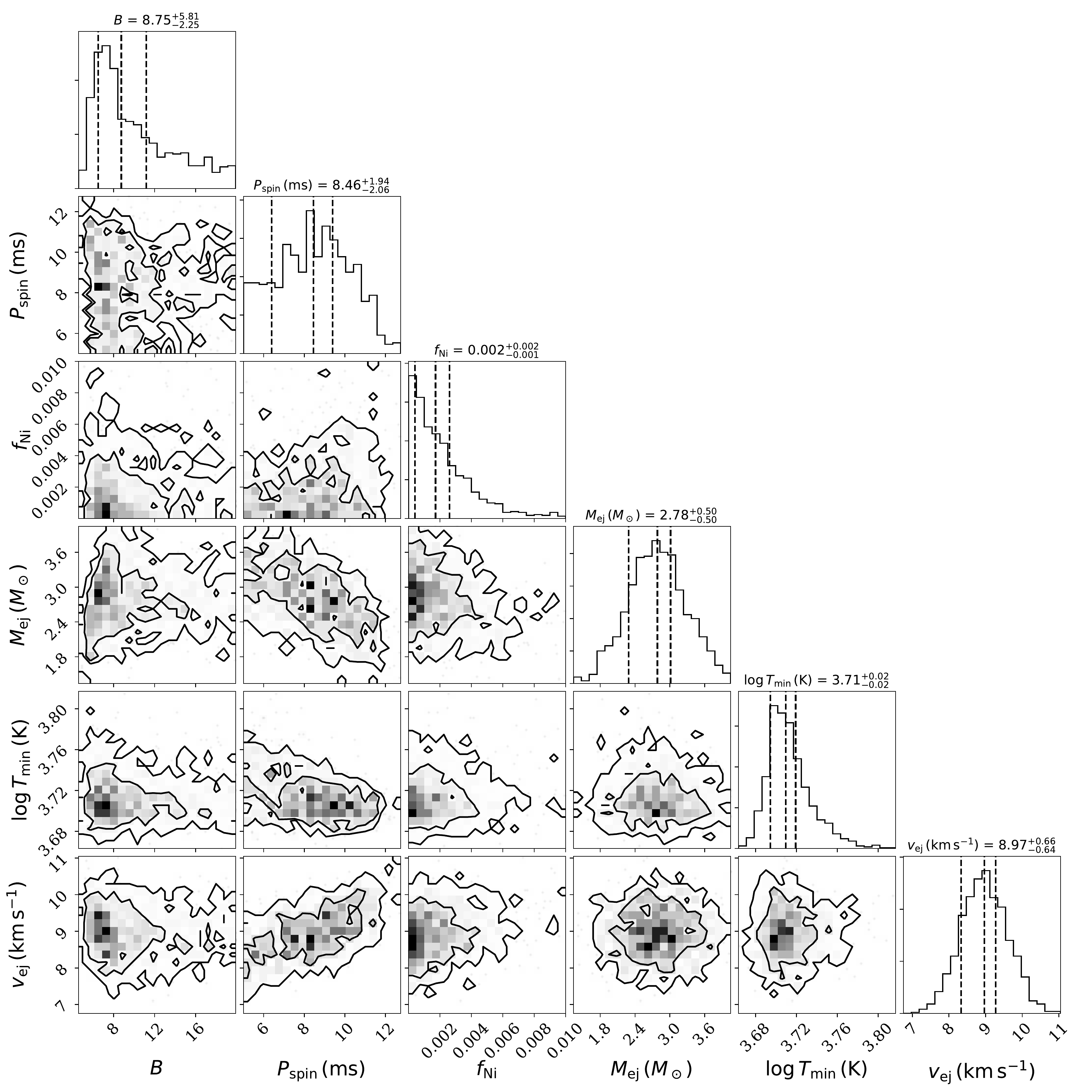}
\caption{Corner plot of the combined $^{56}$Ni decay and magnetar model fit to photometry of DES14X2fna using MOSFiT, showing the best-fit parameter values and their correlations and degeneracies. Dashed lines represent 16th, 50th and 84th percentiles and contours represent 1$\sigma$ and 2$\sigma$ confidence levels.}
\label{magni_corner}
\end{minipage}
\end{figure*}

Figure \ref{magni_corner} shows the corner plot detailing best-fit parameter values and their correlations with each other for the combined $^{56}$Ni decay and CSM interaction model.

\section*{Acknowledgements}

We thank the anonymous referee for the comments and suggestions that have helped to improve the paper.

M.G., C.P.G. and M.S. acknowledge support from EU/FP7-ERC grant No. [615929].

This work was supported by the Science and Technology Facilities Council [grant number ST/P006760/1] through the DISCnet Centre for Doctoral Training. MS acknowledges support from EU/FP7-ERC grant 615929.

L.G. was funded by the European Union's Horizon 2020 research and innovation programme under the Marie Sk\l{}odowska-Curie grant agreement No. 839090. This work has been partially supported by the Spanish grant PGC2018-095317-B-C21 within the European Funds for Regional Development (FEDER).

Funding for the DES Projects has been provided by the U.S. Department of Energy, the U.S. National Science Foundation, the Ministry of Science and Education of Spain, the Science and Technology Facilities Council of the United Kingdom, the Higher Education Funding Council for England, the National Center for Supercomputing Applications at the University of Illinois at Urbana-Champaign, the Kavli Institute of Cosmological Physics at the University of Chicago, the Center for Cosmology and Astro-Particle Physics at the Ohio State University, the Mitchell Institute for Fundamental Physics and Astronomy at Texas A\&M University, Financiadora de Estudos e Projetos, Funda{\c c}{\~a}o Carlos Chagas Filho de Amparo {\`a} Pesquisa do Estado do Rio de Janeiro, Conselho Nacional de Desenvolvimento Cient{\'i}fico e Tecnol{\'o}gico and
the Minist{\'e}rio da Ci{\^e}ncia, Tecnologia e Inova{\c c}{\~a}o, the Deutsche Forschungsgemeinschaft and the Collaborating Institutions in the Dark Energy Survey. 

The Collaborating Institutions are Argonne National Laboratory, the University of California at Santa Cruz, the University of Cambridge, Centro de Investigaciones Energ{\'e}ticas, Medioambientales y Tecnol{\'o}gicas-Madrid, the University of Chicago, University College London, the DES-Brazil Consortium, the University of Edinburgh, the Eidgen{\"o}ssische Technische Hochschule (ETH) Z{\"u}rich, Fermi National Accelerator Laboratory, the University of Illinois at Urbana-Champaign, the Institut de Ci{\`e}ncies de l'Espai (IEEC/CSIC), the Institut de F{\'i}sica d'Altes Energies, Lawrence Berkeley National Laboratory, the Ludwig-Maximilians Universit{\"a}t M{\"u}nchen and the associated Excellence Cluster Universe,
the University of Michigan, the National Optical Astronomy Observatory, the University of Nottingham, The Ohio State University, the University of Pennsylvania, the University of Portsmouth, SLAC National Accelerator Laboratory, Stanford University, the University of Sussex, Texas A\&M University, and the OzDES Membership Consortium.

Based in part on observations at Cerro Tololo Inter-American Observatory, National Optical Astronomy Observatory, which is operated by the Association of
Universities for Research in Astronomy (AURA) under a cooperative agreement with the National Science Foundation.

The DES data management system is supported by the National Science Foundation under Grant Numbers AST-1138766 and AST-1536171. The DES participants from Spanish institutions are partially supported by MINECO under grants AYA2015-71825, ESP2015-66861, FPA2015-68048, SEV-2016-0588, SEV-2016-0597, and MDM-2015-0509, some of which include ERDF funds from the European Union. IFAE is partially funded by the CERCA program of the Generalitat de Catalunya.
Research leading to these results has received funding from the European Research
Council under the European Union's Seventh Framework Program (FP7/2007-2013) including ERC grant agreements 240672, 291329, and 306478. We  acknowledge support from the Brazilian Instituto Nacional de Ci\^encia e Tecnologia (INCT) e-Universe (CNPq grant 465376/2014-2).

This manuscript has been authored by Fermi Research Alliance, LLC under Contract No. DE-AC02-07CH11359 with the U.S. Department of Energy, Office of Science, Office of High Energy Physics.

Based in part on data acquired at the Anglo-Australian Telescope, under program A/2013B/012. We acknowledge the traditional owners of the land on which the AAT stands, the Gamilaraay people, and pay our respects to elders past and present.

This manuscript has been authored by Fermi Research Alliance, LLC under Contract No. DE-AC02-07CH11359 with the U.S. Department of Energy, Office of Science, Office of High Energy Physics. The United States Government retains and the publisher, by accepting the article for publication, acknowledges that the United States Government retains a non-exclusive, paid-up, irrevocable, world-wide license to publish or reproduce the published form of this manuscript, or allow others to do so, for United States Government purposes.

\section*{Data Availability Statement}

The data underlying this article regarding DES14X2fna are available in the article and  through the WISeREP (https://wiserep.weizmann.ac.il/home) archive \citep{wiserep}. Comparison data was sourced from published literature with sources detailed in tables \ref{IIb_table} and \ref{Ic_table}.

\bibliographystyle{mnras}
\bibliography{refs}

\section*{Affiliations}

$^{1}$ School of Physics and Astronomy, University of Southampton,  Southampton, SO17 1BJ, UK\\
$^{2}$ DISCnet Centre for Doctoral Training, University of Southampton, Southampton SO17 1BJ, UK\\
$^{3}$ Institute of Cosmology and Gravitation, University of Portsmouth, Portsmouth, PO1 3FX, UK\\
$^{4}$ CENTRA, Instituto Superior T\'ecnico, Universidade de Lisboa, Av. Rovisco Pais 1, 1049-001 Lisboa, Portugal\\
$^{5}$ The Research School of Astronomy and Astrophysics, Australian National University, ACT 2601, Australia\\
$^{6}$ Departamento de F\'isica Te\'orica y del Cosmos, Universidad de Granada, E-18071 Granada, Spain\\
$^{7}$ INAF, Astrophysical Observatory of Turin, I-10025 Pino Torinese, Italy\\
$^{8}$ Centre for Astrophysics \& Supercomputing, Swinburne University of Technology, Victoria 3122, Australia\\
$^{9}$ Sydney Institute for Astronomy, School of Physics, A28, The University of Sydney, NSW 2006, Australia\\
$^{10}$ Universit\'e Clermont Auvergne, CNRS/IN2P3, LPC, F-63000 Clermont-Ferrand, France\\
$^{11}$ School of Mathematics and Physics, University of Queensland,  Brisbane, QLD 4072, Australia\\
$^{12}$McDonald Observatory, The University of Texas at Austin, Fort Davis, TX 79734\\
$^{13}$ Cerro Tololo Inter-American Observatory, NSF's National Optical-Infrared Astronomy Research Laboratory, Casilla 603, La Serena, Chile\\
$^{14}$ Departamento de F\'isica Matem\'atica, Instituto de F\'isica, Universidade de S\~ao Paulo, CP 66318, S\~ao Paulo, SP, 05314-970, Brazil\\
$^{15}$ Laborat\'orio Interinstitucional de e-Astronomia - LIneA, Rua Gal. Jos\'e Cristino 77, Rio de Janeiro, RJ - 20921-400, Brazil\\
$^{16}$ Instituto de Fisica Teorica UAM/CSIC, Universidad Autonoma de Madrid, 28049 Madrid, Spain\\
$^{17}$ CNRS, UMR 7095, Institut d'Astrophysique de Paris, F-75014, Paris, France\\
$^{18}$ Sorbonne Universit\'es, UPMC Univ Paris 06, UMR 7095, Institut d'Astrophysique de Paris, F-75014, Paris, France\\
$^{19}$ Department of Physics and Astronomy, Pevensey Building, University of Sussex, Brighton, BN1 9QH, UK\\
$^{20}$ Department of Physics \& Astronomy, University College London, Gower Street, London, WC1E 6BT, UK\\
$^{21}$ Instituto de Astrofisica de Canarias, E-38205 La Laguna, Tenerife, Spain\\
$^{22}$ Universidad de La Laguna, Dpto. Astrofísica, E-38206 La Laguna, Tenerife, Spain\\
$^{23}$ Department of Astronomy, University of Illinois at Urbana-Champaign, 1002 W. Green Street, Urbana, IL 61801, USA\\
$^{24}$ National Center for Supercomputing Applications, 1205 West Clark St., Urbana, IL 61801, USA\\
$^{25}$ Institut de F\'{\i}sica d'Altes Energies (IFAE), The Barcelona Institute of Science and Technology, Campus UAB, 08193 Bellaterra (Barcelona) Spain\\
$^{26}$ INAF-Osservatorio Astronomico di Trieste, via G. B. Tiepolo 11, I-34143 Trieste, Italy\\
$^{27}$ Institute for Fundamental Physics of the Universe, Via Beirut 2, 34014 Trieste, Italy\\
$^{28}$ Observat\'orio Nacional, Rua Gal. Jos\'e Cristino 77, Rio de Janeiro, RJ - 20921-400, Brazil\\
$^{29}$ Centro de Investigaciones Energ\'eticas, Medioambientales y Tecnol\'ogicas (CIEMAT), Madrid, Spain\\
$^{30}$ Department of Physics, IIT Hyderabad, Kandi, Telangana 502285, India\\
$^{31}$ Fermi National Accelerator Laboratory, P. O. Box 500, Batavia, IL 60510, USA\\
$^{32}$ Santa Cruz Institute for Particle Physics, Santa Cruz, CA 95064, USA\\
$^{33}$ Institute of Theoretical Astrophysics, University of Oslo. P.O. Box 1029 Blindern, NO-0315 Oslo, Norway\\
$^{34}$ Institut d'Estudis Espacials de Catalunya (IEEC), 08034 Barcelona, Spain\\
$^{35}$ Institute of Space Sciences (ICE, CSIC),  Campus UAB, Carrer de Can Magrans, s/n,  08193 Barcelona, Spain\\
$^{36}$ Kavli Institute for Cosmological Physics, University of Chicago, Chicago, IL 60637, USA\\
$^{37}$ Department of Physics, Stanford University, 382 Via Pueblo Mall, Stanford, CA 94305, USA\\
$^{38}$ Kavli Institute for Particle Astrophysics \& Cosmology, P. O. Box 2450, Stanford University, Stanford, CA 94305, USA\\
$^{39}$ SLAC National Accelerator Laboratory, Menlo Park, CA 94025, USA\\
$^{40}$ Faculty of Physics, Ludwig-Maximilians-Universit\"at, Scheinerstr. 1, 81679 Munich, Germany\\
$^{41}$ Max Planck Institute for Extraterrestrial Physics, Giessenbachstrasse, 85748 Garching, Germany\\
$^{42}$ Universit\"ats-Sternwarte, Fakult\"at f\"ur Physik, Ludwig-Maximilians Universit\"at M\"unchen, Scheinerstr. 1, 81679 M\"unchen, Germany\\
$^{43}$ Australian Astronomical Optics, Macquarie University, North Ryde, NSW 2113, Australia\\
$^{44}$ Lowell Observatory, 1400 Mars Hill Rd, Flagstaff, AZ 86001, USA\\
$^{45}$ Center for Cosmology and Astro-Particle Physics, The Ohio State University, Columbus, OH 43210, USA\\
$^{46}$ Department of Physics, The Ohio State University, Columbus, OH 43210, USA\\
$^{47}$ George P. and Cynthia Woods Mitchell Institute for Fundamental Physics and Astronomy, and Department of Physics and Astronomy, Texas A\&M University, College Station, TX 77843,  USA\\
$^{48}$ Department of Astronomy, The Ohio State University, Columbus, OH 43210, USA\\
$^{49}$ Radcliffe Institute for Advanced Study, Harvard University, Cambridge, MA 02138\\
$^{50}$ Instituci\'o Catalana de Recerca i Estudis Avan\c{c}ats, E-08010 Barcelona, Spain\\
$^{51}$ Physics Department, 2320 Chamberlin Hall, University of Wisconsin-Madison, 1150 University Avenue Madison, WI  53706-1390\\
$^{52}$ Institute of Astronomy, University of Cambridge, Madingley Road, Cambridge CB3 0HA, UK\\
$^{53}$ Department of Astrophysical Sciences, Princeton University, Peyton Hall, Princeton, NJ 08544, USA\\
$^{54}$ Department of Physics and Astronomy, University of Pennsylvania, Philadelphia, PA 19104, USA\\
$^{55}$ Department of Physics, University of Michigan, Ann Arbor, MI 48109, USA\\
$^{56}$ Computer Science and Mathematics Division, Oak Ridge National Laboratory, Oak Ridge, TN 37831\\
\bsp	
\label{lastpage}
\end{document}